\documentclass[apjl]{emulateapj}
\usepackage{rotating}
\usepackage{threeparttable}
\usepackage{color}
\usepackage{graphicx}
\usepackage{datetime}
\def\rpm {~$\frac{rad}{m^2}$~}

\begin{document}
\title{The distribution of polarized radio sources $>$15$\mu$Jy in GOODS-N}
\shorttitle{GOODS polarization}
\author{L. Rudnick\altaffilmark{1}, F. N. Owen \altaffilmark{2}}


\altaffiltext{1}{Minnesota Institute for Astrophysics, School of Physics and Astronomy, University of Minnesota, 116 Church 
Street SE, Minneapolis, MN  55455;  corresponding author: larry@umn.edu}
\altaffiltext{2}{National Radio Astronomy Observatory, Socorro, NM 87801}

\begin{abstract} We present deep VLA observations of the polarization of radio sources in the GOODS-N field at 1.4~GHz at 
resolutions of 1.6\arcsec and 10\arcsec. At 1.6\arcsec, we find that
the peak flux cumulative number count distribution is N($>$p) $\sim$  45 * (p/30$\mu$Jy)$^{-0.6}$  per square degree above a detection threshold of 14.5~$\mu$Jy. 
This represents a break from the steeper slopes at higher flux densities, resulting in fewer sources predicted for future surveys with the 
SKA and its precursors. It provides a significant challenge for using background RMs to study clusters of galaxies or individual galaxies.  Most of the polarized sources are well above our detection limit, and
are radio galaxies which are well-resolved even at 10\arcsec, with redshifts from $\sim$0.2 - 1.9.  We determined a total polarized flux for each source by integrating the 10\arcsec~polarized intensity maps, as will be done by upcoming surveys such as POSSUM. These total polarized fluxes are a factor of 2 higher, on average, than the peak polarized flux at 1.6\arcsec; this would increase the number counts by $\sim$50\% at a fixed flux level. The detected sources have rotation measures (RMs) with a characteristic rms scatter of $\sim$11$\frac{rad}{m^2}$ around the local
Galactic value, after eliminating likely outliers. The median fractional polarization from all total intensity sources does not continue the trend of increasing at lower flux densities, as seen for stronger sources.  The changes in the polarization characteristics seen at these low fluxes likely represent the increasing dominance of star-forming galaxies. 

\end{abstract}

\keywords{Galaxies:active --- Polarization}

\clearpage
\section{INTRODUCTION}

Deep surveys for polarized radio sources are motivated by several different scientific goals.   One is to use their rotation measures (RMs) as a background grid to study the magneto-ionic structure of the Milky Way \citep{taylornvss}, or nearby galaxies \citep{step08}, or clusters of galaxies \citep{govoni10}, and in the unique case of Centaurus~A, even the lobes of radio galaxies \citep{fornax}.   The most extensive RM grid currently available is that of \cite{taylornvss}, with 37,543 polarized sources reconstructed from the NVSS \citep{NVSS} survey.   It has a source density of $\sim$1 per square degree, making it an excellent probe of the large scale structure of the Milky Way, but insufficient for studies of even nearby extended extragalactic objects.   This has motivated ambitious plans for polarization surveys such as POSSUM \citep{possum1} on the Australian SKA precursor (ASKAP, \cite{askap} ), with a target of polarized source densities $\sim$100 times higher than the NVSS.

A second motivation for deep polarization surveys is to probe source populations, e.g., as a function of flux level, luminosity and redshift.  Polarization counts offer complementary information to counts in total intensity (e.g., \cite{icounts,morrisonG}), and can help distinguish between the contributions of star-forming galaxies, which are expected to have low polarizations, from AGN radio sources.

Finally, polarization surveys can inform the physical models of radio galaxies, e.g., the amount of magnetic field order/disorder,  the presence of an external magnetised medium and even the interactions  of the thermal and relativistic plasmas \citep{guidetti12}.  For these purposes, the RMs, the depolarization as a function of wavelength, and eventually the spatially unresolved Faraday structure, e.g. \cite{law}, are important diagnostics.

The deepest polarization surveys to date are those of the ELAIS-N1 field by \cite{elais2}, which reach down to polarized fluxes p$>$270~$\mu$Jy, and the ATLBS field studied by \cite{atlbs}.\footnote{ As discussed below, we consider the latter work reliable only for p$>$1000~$\mu$Jy.}  The very sensitive 1.4~GHz Karl G. Jansky Very Large Array (VLA) survey of the GOODS-N field \citep{morrisonG}, with an rms sensitivity of several $\mu$Jy, provided an excellent opportunity to extend polarization studies to much lower levels, albeit over a smaller area of the sky.   A parallel effort to reach polarized sensitivity levels in the $\mu$Jy range is based on the stacking of sources from the NVSS \citep{stack}, and is complementary to the work presented here.

This paper is arranged as follows.   Section II describes the observations and data reduction, while Section III describes the methodology behind our source detection and presents the results. More detailed information on individual sources is presented in Section IV.   In Section V, we derive the polarized number counts and fractional polarization distributions as a function of flux density, and compare these to results at higher fluxes and models seeking to predict the behavior at $\mu$Jy levels.  Section VI contains a discussion of the results, in particular their implications for some key experiments planned for the next generation of radio telescopes.

\section{OBSERVATIONS \& DATA REDUCTION}

\subsection{Observations and initial reduction}

	The GOODS-N field was observed for a total of 39 hours including calibration and move time between
August 9 and September 11, 2011. There was a single pointing centered at RA: 12h36m49.4s, Dec: 62$^o$12'58'' . Eight different scheduling blocks were observed, seven of 5 hours, and one of 4 hours.
Approximately 33 hours of this time were spent on-source. The observations covered
the bands from $1000-1512$ and $1520-2032$ MHz using 1 MHz channels. A phase, bandpass and  instrumental polarization
calibrator, J1313+6735, was observed every twenty minutes. 3C286 was observed to calibrate the flux density scale
and the polarization position angle. 

	For each scheduling block, the data were edited and calibrated in AIPS. The worst parts of the band, in 
particular between 1520 and 1648 MHz were flagged at the beginning of this process. The rest of the dataset was edited 
using the RFLAG task. After total intensity calibration, the uv-data weights were calibrated using the AIPS task, 
REWAY. More details on the total intensity calibration are contained in \cite{owen14}. 

	The total intensity data were imaged in CASA using the wide-field, Multi-Scale-Multi-Frequency-Synthesis algorithm (MSMFS, \cite{rau}) with nterms=2. This imaging algorithm solves for the total intensity and spectral index image across the full 
bandwidth --in this case,  1-2 GHz. The maps were reconstructed with a  $1.6\arcsec\times1.6\arcsec$ FWHM Gaussian beam, and then also later convolved to $10\arcsec\times10\arcsec$ for extended sources. 
More details are given in \cite{owen14}. For this analysis, the primary beam correction was not applied to the initial maps, in order to keep the noise properties relatively constant across the field.  All total intensities reported in this paper were later corrected for the primary beam using the CASA task {\em widebandpbcor}.

\subsection{Polarization processing}

\subsubsection{Polarization calibration and imaging}
The polarization data were also calibrated in AIPS. First, a correction was applied for the ionospheric Faraday
Rotation using the AIPS procedure VLATECR, which uses an on-line, time variable ionospheric model supplied by
JPL. For these data the corrections were very small, only a few ~\rpm~ and show that the ionosphere was not
active during the observations. The $R-L$ delay was then calculated using RLDLY. Next PCAL was used with the
observations of J1313+6735 to estimate the instrumental polarization as a function of frequency for each observed channel. These
calibrations showed that, as expected, the instrumental polarization rises rapidly at the low end of the band,
especially below 1100 MHz.  Finally the polarization position angle was calibrated using 3C286. 

To estimate the residual instrumental polarization after correction, we looked at the seven strongest total intensity sources, i.e., those above 3~mJy, using the full 1.6\arcsec~ resolution.  Six of these showed strong polarization above our detection threshold of 14.5~$\mu$Jy, corresponding to fractional polarizations from 0.4\% to 8\%, as reported below.   Even the lowest one is likely real, since it has an RM=19~\rpm (the Galactic foreground value) whereas 0~\rpm~ would be expected from instrumental residuals. The one remaining strong source had  I=3964~$\mu$Jy and p=9~$\mu$Jy (consistent with noise), corresponding to a fractional polarization of 0.2\%.   We therefore consider our residual instrumental polarization to be $<$~0.4\%.  

	The polarization Q and U image cubes were made in AIPS with IMAGR, using frequency planes averaged over 10~MHz. For cleaning, 168 512x512 facets were used to cover the useful field-of-view. The data at frequencies lower than 1250 MHz were 
not included due to higher noise, high instrumental polarization and interference issues. After the uv-data flagging 
described earlier, the final Q and U cubes have a total of 60 10~MHz image planes with center frequencies between 1265 MHz 
and 2027 MHz. Each plane in the cubes was then convolved to a common resolution of $1.6\arcsec\times1.6\arcsec$ Gaussian FWHM. 
We also convolved each plane of the Q,U cubes to 10\arcsec~ to better recover the flux of extended sources, as described further below. 

In order to combine the Q,U data across the entire bandwidth, we have to first remove their total intensity spectral dependence\footnote{This normalization is required by Faraday synthesis \citep{brentjens} so that, after correction, Q($\lambda^2$), U($\lambda^2$) are constant amplitude sinusoidal waves in the case of a simple Faraday screen.}. Ideally, this would be done on a source by source basis.  However, we did not have an adequate characterization of the errors in the spectra from MSMFS in its early development, when the analysis was carried out.   We therefore applied a nominal correction for the spectral slope as a function of distance from the field center, which accounts for both the average intrinsic source spectral slope and the steepening due to the change in primary beam size as a function of frequency. Q,U cutout cubes for each source were thus normalized to a nominal frequency of 1500~MHz. After determining the polarized fluxes from Faraday synthesis, as described in Section \ref{faraday}, we then applied a primary beam correction appropriate to a monochromatic beam at 1500~MHz.

This empirical spectral correction does not account for spectral curvature that occurs due to the primary beams, especially at large distances, nor does it account for real variations in the intrinsic source spectra.  We did not expect these to significantly bias our results, especially since the weighting of the spectral channels for the polarization analysis (described below) gave us an effective bandwidth much narrower than the full total intensity observations. In particular, the spectral corrections due to curvature would be largest at the high frequencies, but as shown below, these are weighted down quite strongly in the Faraday analysis.  However, to confirm our expectations, we performed a series of experiments by taking data from strongly polarized sources at $\sim$1000'' from field center.  We normalized their Q,U data using a wide range of spectral slopes (e.g., -4.1 to -1.1, bracketing the slope of $\sim$-2.6 actually used at 1000\arcsec), and then ran each trial through our full analysis.  The changes in polarized flux were $\pm$3\% over this very wide spectral range, so they do not significantly affect our ability to detect sources, despite the possibility of spectral errors up to $\pm$1. However, for the one detected source where the signal:noise is $>$30 in polarized flux, the spectral uncertainty would increase the error over what we report.

 \begin{figure*}[!ht]
\begin{center}
\includegraphics[width=6.5cm]{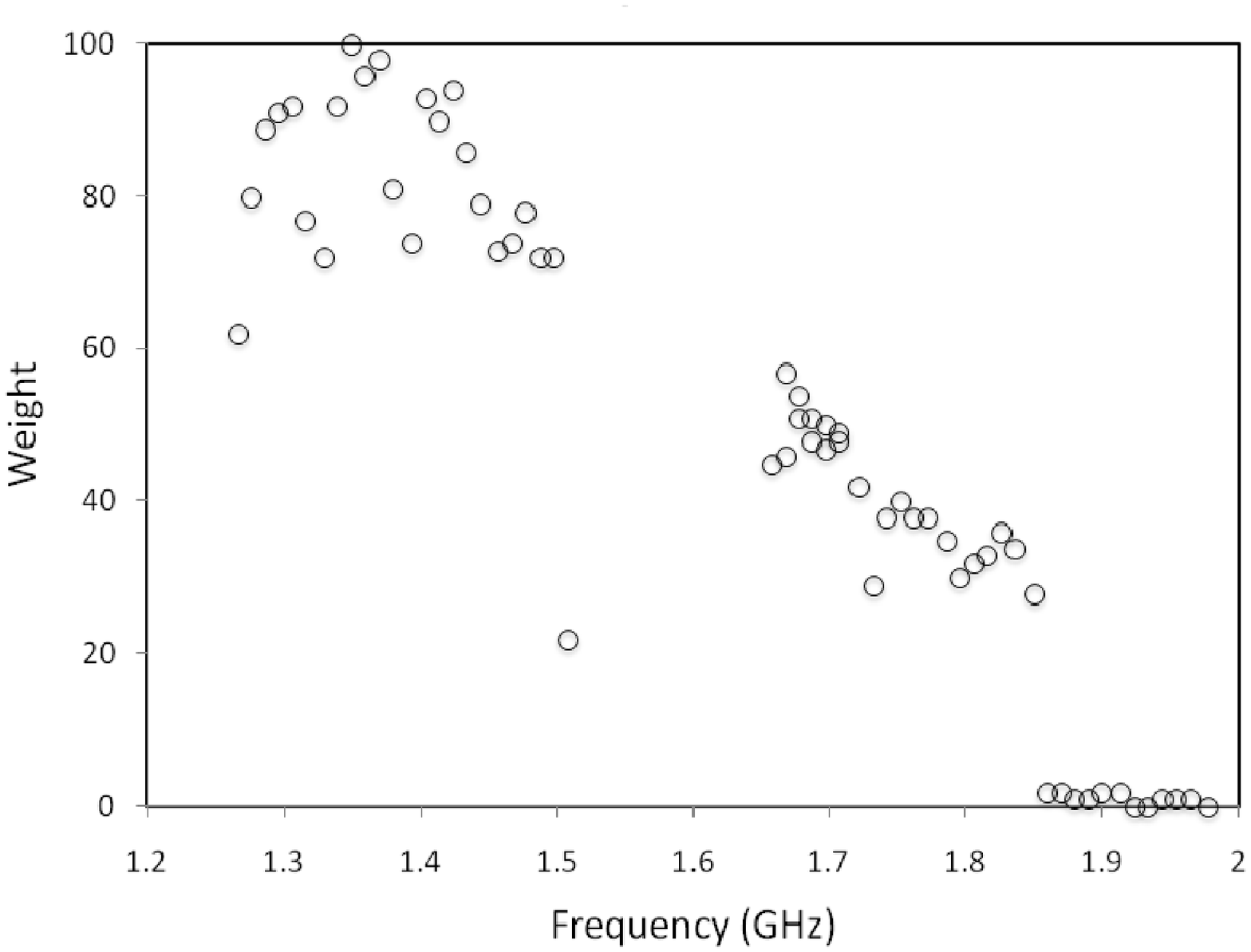} \hskip 0.25in
\includegraphics[width=6cm]{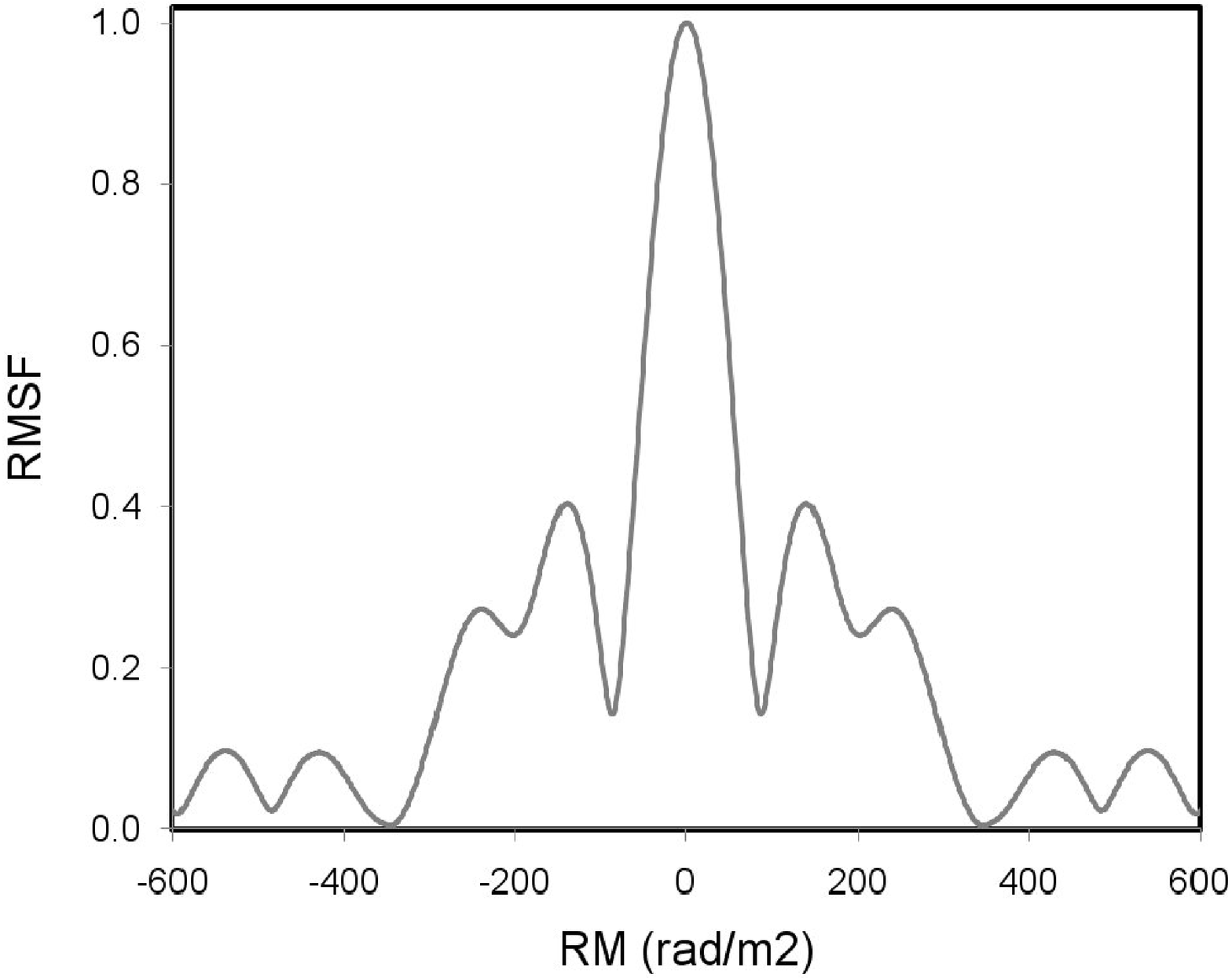}
\end{center}
\caption{\footnotesize{Left: Weights ( W($\lambda^2$) ) used in calculation of $\vec{\Phi}(RM,x,y)$. Right: Rotation measure structure function.}}
\label{rmsf}
\end{figure*}

\subsubsection{Faraday synthesis}
\label{faraday}
After spectral normalization, in order to achieve maximum sensitivity, we need to average Q and U over the entire frequency band.  However, given the large bandwidths in these observations, polarized sources with high rotation measures would be depolarized with such a simple averaging.  We therefore used the Faraday Synthesis technique \citep{brentjens}, as implemented in the AIPS task FARS to construct a Faraday spectrum cube $\vec{\Phi}(RM,x,y)$ for each source, with -600~\rpm~$<$~RM~$<$~+600~\rpm and 4 \rpm~bins.  This is essentially the Fourier transform  
$\vec{\Phi}(RM,x,y)$=$\int{\vec{P}(x,y,\lambda^2)W(\lambda^2)e^{2{\pi}iRM\lambda^2}}d\lambda^2$
where $\vec{P}(\lambda^2) = Q(\lambda^2) + iU(\lambda^2)$ and $W(\lambda^2)$ is a weighting function.  

The choice of a weighting function is a compromise between the best signal-to-noise, where $W(\lambda^2)$ is inversely proportional to the variance in each frequency (wavelength) channel, and a uniform weighting, which gives the highest resolution in Faraday space, or even a higher weighting at high frequencies to compensate for possible depolarization at longer wavelengths.  We experimented with different weightings, and they produced only modest changes in the noise properties and source strengths.  The final weighting of individual channels, including the elimination of some high frequency channels with very high noise, is shown in Figure \ref{rmsf}. The final noise-equivalent bandwidth was $\sim$~300~MHz, with minimum (maximum) frequencies of 1.2625 (1.8965) GHz.  The FWHM for the peak in the Faraday response function (RMSF) was 106 \rpm, (see Fig. \ref{rmsf}) with a mean(rms) in each $|\Phi(RM)|$ image of $\sim$10(1.8)$\mu$Jy/beam/RMSF.  Because of the non-linear response at low signal:noise, we empirically determined the  errors in polarized flux by inserting fake sources into the Q,U data, and found an rms scatter of $\sim$3.3~$\mu$Jy around the correct value.  

For every spatial pixel (x,y)  in the image, we then used the AIPS task AFARS to search for the maximum in the Faraday spectrum amplitude $\Phi(RM,x,y) \equiv |\vec{\Phi}(RM, x,y)|$, and created two output images/data sets,  $\Phi_{max}(x,y)$ and $RM_{max}(x,y)$, where $\Phi_{max}(x,y)$ is the maximum amplitude over all RMs ($\pm$600\rpm) at pixel {\em x,y} and  $RM_{max}$ is the rotation measure at which the peak amplitude for that pixel occurs .  Both the maximum amplitude and RM were interpolated between RM bins by AFARS. This is the most direct counterpart to measurements in the literature which previously measured Q and U in two or more well-separated bands and fit for P($\lambda^2$=0) and RM.
 We also produced the corresponding $\Phi_{max10}(x,y)$ and $RM_{max10}$(x,y) images at 10\arcsec~ resolution, through processing of the 10\arcsec~ Q,U cubes. For the rest of the paper, for simplicity, we will simply refer to $\Phi_{max}$ and $RM_{max}$ as {\em P} and {\em RM} for each source, or {\em P$_{10}$} and {\em RM$_{10}$} for the 10\arcsec~ results.

  We did not attempt to deconvolve the Faraday spectra to search for multiple Faraday components because of ambiguities in  such deconvolutions, as described in \cite{farns}.   We have conducted additional experiments, reported elsewhere, showing that these reconstruction techniques are not yet reliable for an arbitrary distribution of Faraday depths over an RM  range comparable to the RMSF. To the extent that there is Faraday structure in an individual source, the RMs reported here (as in the literature) represent a weighted mean of the underlying RM distribution.

\section{SOURCE DETECTION}

{\bf Detection Threshold.}  We searched for polarized sources only at the location of sources detected in the full resolution total intensity image, which had an rms noise of 2.4$\mu$Jy/beam.  We used the AIPS task SAD to identify 506 sources with 
I$_{peak} \geq$28~$\mu$Jy/beam. This high threshold was set because of the need to search for polarized flux, for which the signals would be much weaker. As discussed below, our polarization detection threshold of 14.5$\mu$Jy translates to a $\sim$50\%  polarization for the lowest total intensity sources we searched.  Several of the total intensity sources were rejected after visual inspection because of problems such as detection of spurious sidelobes near very strong sources.  We then recorded the value of P at  the position of  each I$_{peak}$, and the corresponding RM at that position.  

\begin{figure}[!ht]
\begin{center}
\includegraphics[width=8cm]{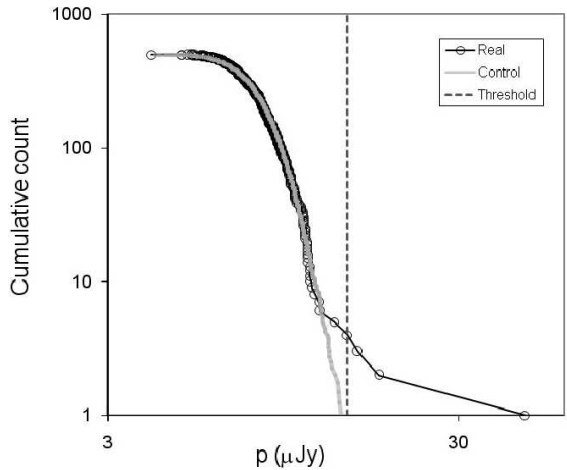}
\end{center}
\caption{\footnotesize{Cumulative distribution of sources as a function of polarized flux. Black: Counts of sources with P$>p$, for total intensity sources with
 I$>$~28~mJy; Grey: Same plot for
control fields. Dotted line shows the adopted detection threshold of 14.5~$\mu$Jy.}}
\label{control}
\end{figure}

  In order to set a detection threshold, we cannot simply use the rms in $\Phi(RM)$; $\Phi_{max}$ is the result of a search over many Faraday depths, and the noise is non-Gaussian.  We therefore determined the detection threshold empirically, using a control sample.  As  controls, we recorded the value of P at the four locations offset by  $\pm$4.5\arcsec~ in each of RA and Dec, i.e., separated from the position of I$_{peak}$ by 6.4\arcsec (4 beams). The peak P for each of the control locations was based on independently searching the entire RM range. We then calculated cumulative histograms of the number of sources (sample and control) as a function of polarized flux level $p$, {\em i.e.,  N(P$>p$)} and plotted these in Figure \ref{control}, after dividing the control counts by 4.

Based on these tests, we adopted a detection threshold of P$>$14.5 $\mu$Jy.  This results in an expected  spurious source detection of only $\frac{1}{4}$ for our 496 remaining sources.  We detected 13 polarized sources above this threshold; 9 of these were extended on scales $>$10\arcsec.  An additional source was detected only by visual examination of the 10\arcsec~ images.

{\bf Completeness and Bias.}  We inserted a series of {polarized test point sources  into the data in order to assess the completeness at different flux levels and to quantify any biases present.  One experiment used the complete processing pipeline starting with the uv data, but was limited in scope because of the practical difficulties in dealing with our 750~Gbyte data set.  We inserted 25 sources with P$_{in}$=20$\mu$Jy and RM=0 (Q=16, U=12) into the multi-channel uv data after calibration, and ran it through the full mapping, deconvolution and Faraday synthesis process, as described above.  We then measured the observed polarized flux at each of the 25 positions, which yielded an average (rms) flux of 20.4 (4)~$\mu$Jy.   Out of the 25 sources, one fell below the 14.5$\mu$Jy threshold and would not have been detected.}

\begin{figure*}[!ht]
\begin{center}
\includegraphics[width=4.2cm]{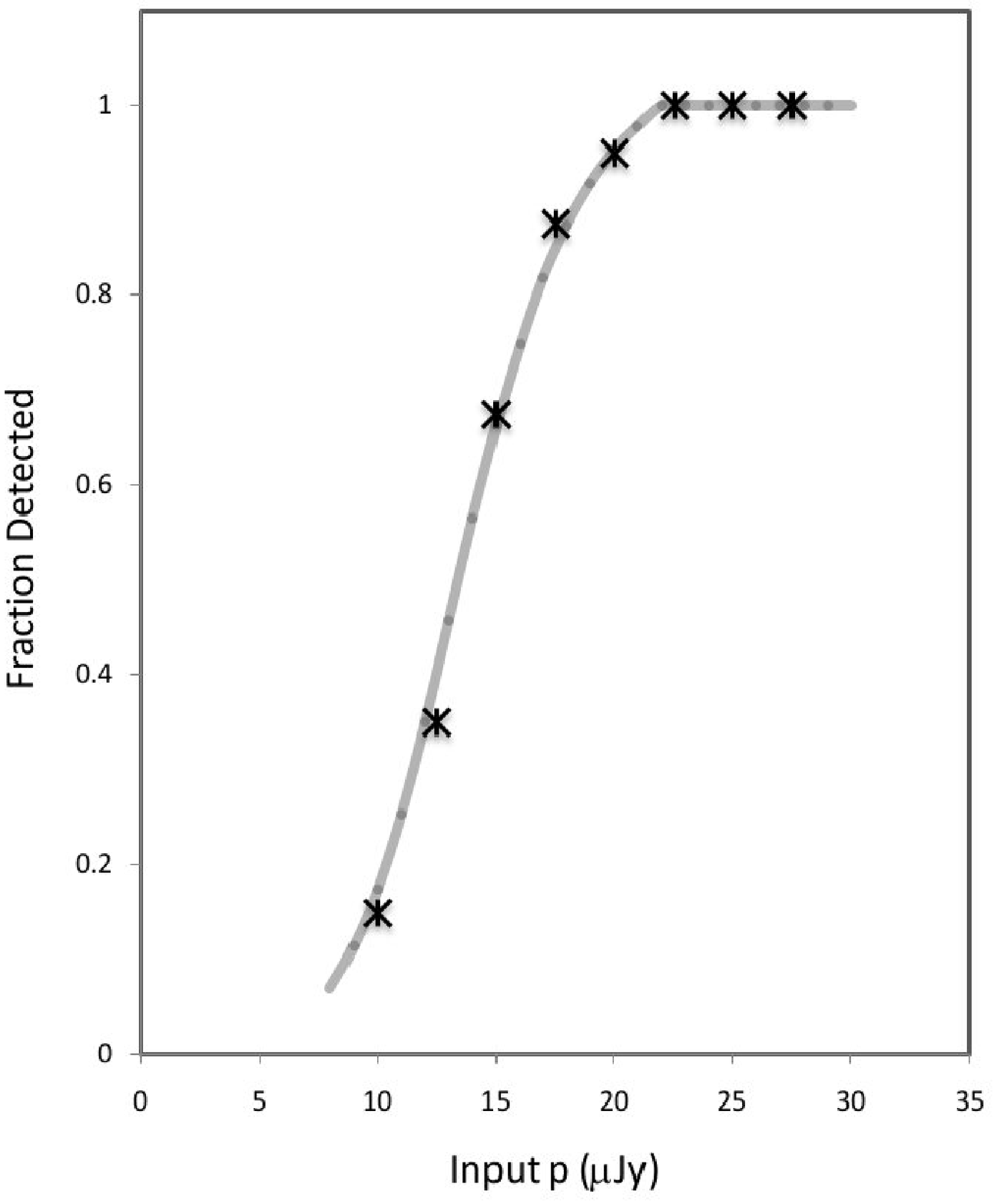}
\includegraphics[width=5cm]{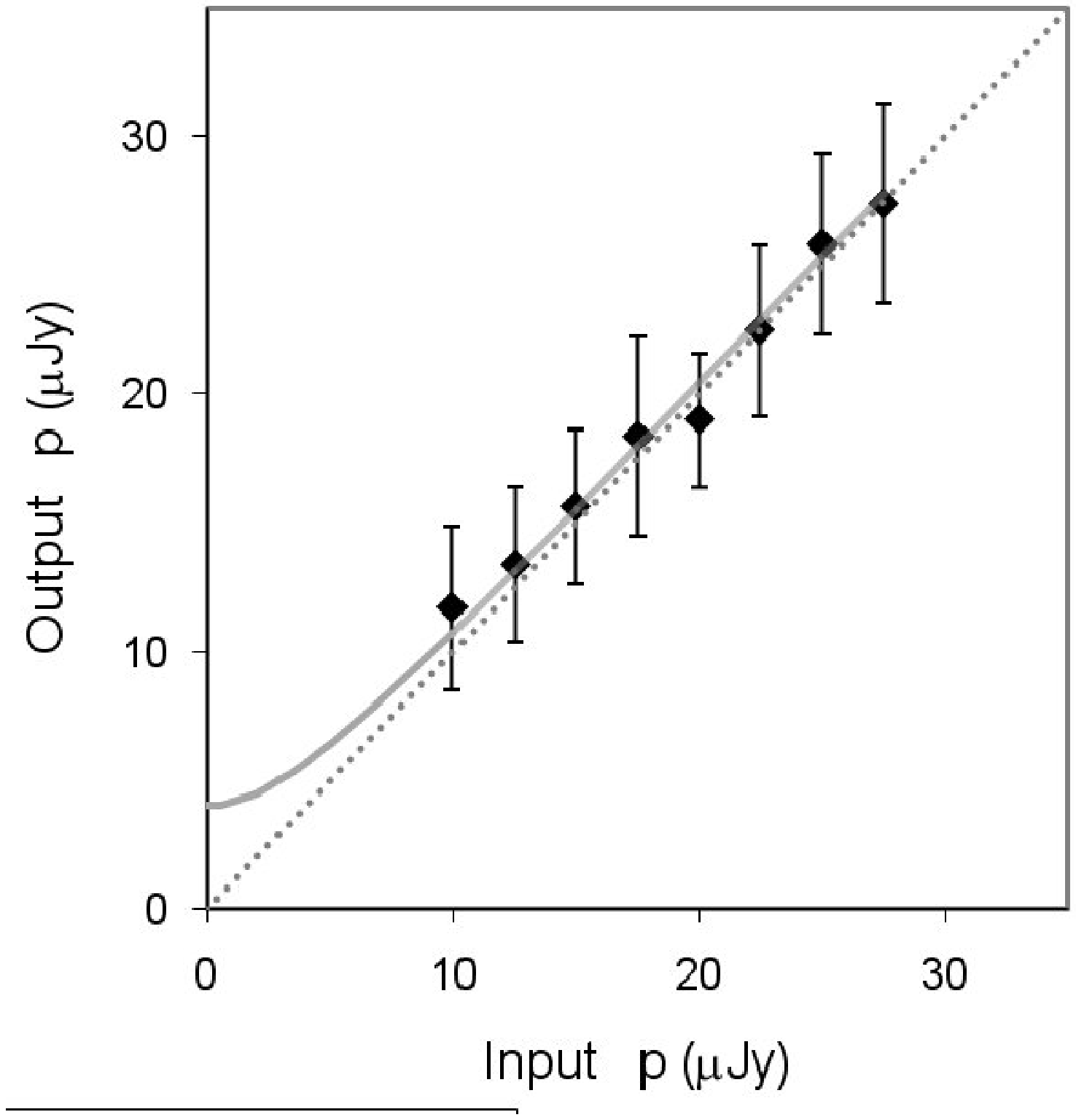} 
\includegraphics[width=4.5cm]{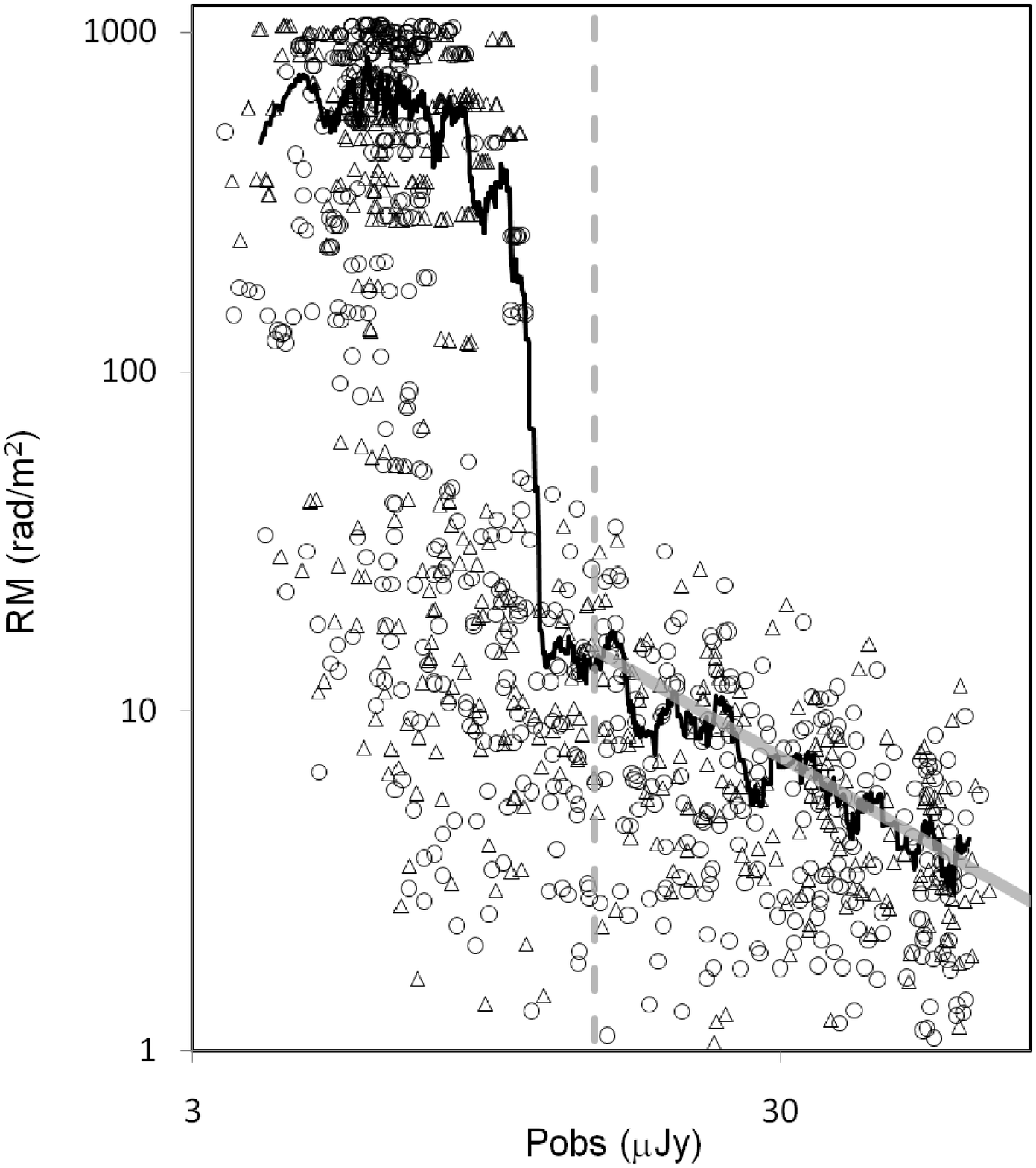}
\end{center}
\label{models}
\caption{\footnotesize{Left: Fractional detection rate (P$_{observed}>$14.5$\mu$Jy) for fake sources inserted into the GOODS field {Q,U cubes}.  Note the small but finite
probability to detect sources below the threshold because of noise biasing. Grey line is an analytic model of completeness used for estimating number
counts.  Middle: Average and rms scatter of detected flux from fake inserted sources as a function of the input flux. The dotted line shows P$_{observed}$ = P$_{in}$, while the grey line shows the result of adding 4$\mu$Jy in quadrature. Right: Absolute value of observed RM (circles: negative; triangles: positive) as a function of observed polarized flux. The black line is $<$RM$^2>$ over a running average of 25 points. The grey line {through the data} indicates the adopted error estimator,  $\sigma_{RM}$ = [{ 220}/P$_{observed}$] \rpm~.  {The vertical dashed line indicates the 14.5$\mu$Jy detection threshold.}
}}
\end{figure*}

{A more extensive set of tests was performed by inserting test polarization signals in the Q,U cubes with RM=0 (Q=U=$\frac{P_{in}}{\sqrt{2}}~\mu$Jy)  to determine the probability of detecting them above the 14.5~$\mu$Jy threshold.  Five groups of twenty-five} sources were inserted at different (x,y) positions  for each of a series of { 8 different} input fluxes P$_{in}$ between 10~$\mu$Jy and 27~$\mu$Jy in increments of 2.5$\mu$Jy,{ with two additional bins up to 50~$\mu$Jy for the RM tests. At each insert position, we searched for the peak in the Faraday spectrum, searching over the range of $\pm$1050~\rpm~ and recorded both the peak amplitude P$_{obs}$ and the RM at which that peak occurred. The search range is broader than used for the real data, and is conservative in that it can result in additional spurious detections.} The results of this experiment are shown in Figure \ref{models}. Because of the positive-definite noise bias, there is a $\sim$15\% probability of detecting a source as faint as 10~$\mu$Jy above the 14.5~$\mu$Jy threshold.  At an input flux of 20~$\mu$Jy, 95\% of the sources are detected { (consistent with our detection of 24/25 sources in the full uv experiment)}. An analytic approximation to the fractional detection rate $f_{det}$ is shown in the figure; $f_{det} = 0.7*atan(1.6*10^{-5}*P_{in}^{4.2})$.  This  expression is used later in estimating the number counts.

We also looked at the average P$_{observed}$ as a function of P$_{in}$ (Figure \ref{models}).  A least squares fit to the average flux yields P$_{observed}$ =$ \sqrt{P_{in}^2 + 4.0~\mu Jy^2}$.  This correction is less than 5\% for even our weakest sources.

To estimate the errors in RM, we used these same experiments to determine the scatter in RM{, $\sigma_{RM}$}, around the inserted value of RM=0, as a function of the strength of the {\em observed} polarized signal (Figure \ref{models}). { $\sigma_{RM}$ is determined as a function of P$_{obs}$ by calculating a running average of $<$RM$^2>$ over 25 points.  The RM can take values anywhere within the search range of $\pm$1050~\rpm.  Note the dramatic transition in  $\sigma_{RM}$ slightly below the detection threshold.  Above the detection threshold, $\sigma_{RM}$ decreases as 1/P, as expected simply due to the signal:noise; the maximum $|RM|$ above threshold is only 35~\rpm.    This result is relevant in our later discussion of weak sources with high RM values.  

 Below the threshold, the 1/P trend is seen to continue, but another population also starts to appear at high values of $|$RM$|$.  These high values are due to noise peaks in the Faraday spectrum which are increasingly above the injected signal peak;  these noise peaks can occur at any value of RM.  Thus, $\sigma_{RM}$ begins to increase dramatically as P$_{obs}$ goes down, as these noise peaks appear more frequently. } 

We visually identified 22 extended (more than $\sim$2 beams) sources in total intensity, and manually examined the images of $\Phi_{max10}$ for evidence of polarized emission. The rms amplitude in empty regions of $\Phi_{max10}$ images was $\sim$~20-25~$\mu$Jy.  Nine of these extended sources, all detected in $\Phi_{max}$ at 1.6\arcsec, were also detected in $\Phi_{max10}$.  One additional source, J123611.2+622810, was detected only in $\Phi_{max10}$, but not at the full resolution $\Phi_{max}$.  A summary of the polarization properties of the detected sources, at both 1.6\arcsec~ and 10\arcsec~ resolution, is given in Table 1.  One of the value-added products of the POSSUM survey will be the total polarized emission from all the components making up each individual radio source.  We have therefore included an equivalent number in Table 1,  the integral of the polarized flux over each source, mapped at 10\arcsec~ resolution.  Note that this is generally greater than the polarized flux that would be observed with a beam larger than the source size, because of interference between components in the low resolution Q and U images.  




\vskip -.15in
\section{INDIVIDUAL SOURCES}
\vskip -.001in
Table 2 complements the radio data on the detections reported in Table 1;  13 out of 14 sources can be identified with optical counterparts; 12 of these have either photometric or spectroscopic redshifts, ranging from $\sim$0.2 - 1.9.  All four of the compact sources have counterparts in the HDF images \citep{giaval} although only two are visible in SDSS \citep{sdss}. Of the extended sources,  the polarized emission is sometimes associated with the hot spot regions or lobes of double radio galaxies and sometimes with the cores or jets of FRI type sources.  Two sources have morphologies which are unclear;  J123550.6+622757 is a $\sim$50~kpc source inside the envelope of a z=0.5 elliptical galaxy;  J123611.2+622810 is quite unusual since it is well resolved in both polarized and total intensity but has no recognizable morphology or optical identification. One of the FRI cores, J123644.3+621132 has been mapped at higher resolution, both with VLBI \citep{chi} and e-MERLIN \citep{guidetti}, showing a likely core-jet morphology connected with the larger-scale FRI jets.

\begin{figure}[!ht]
\begin{center}
\includegraphics[width=6.0cm]{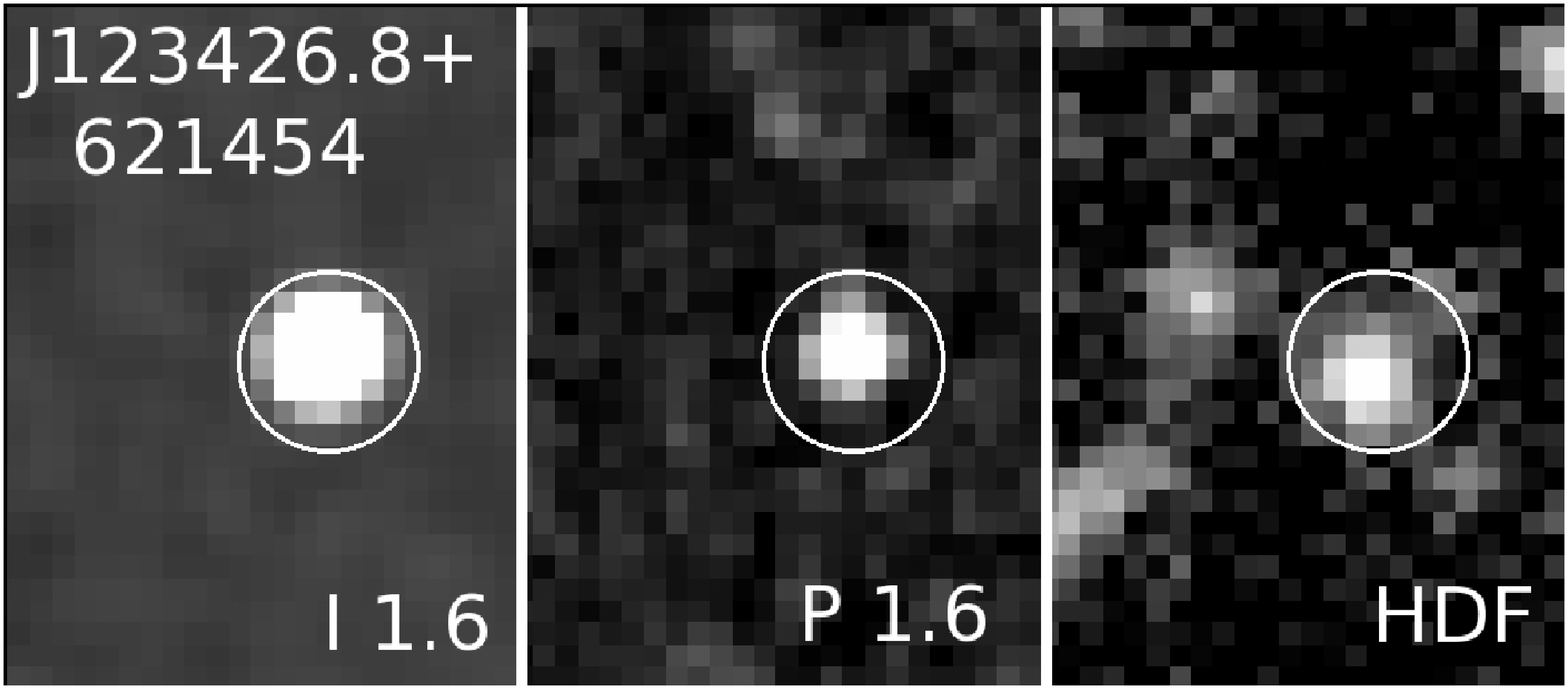}
\vskip .1in
\includegraphics[width=6.0cm]{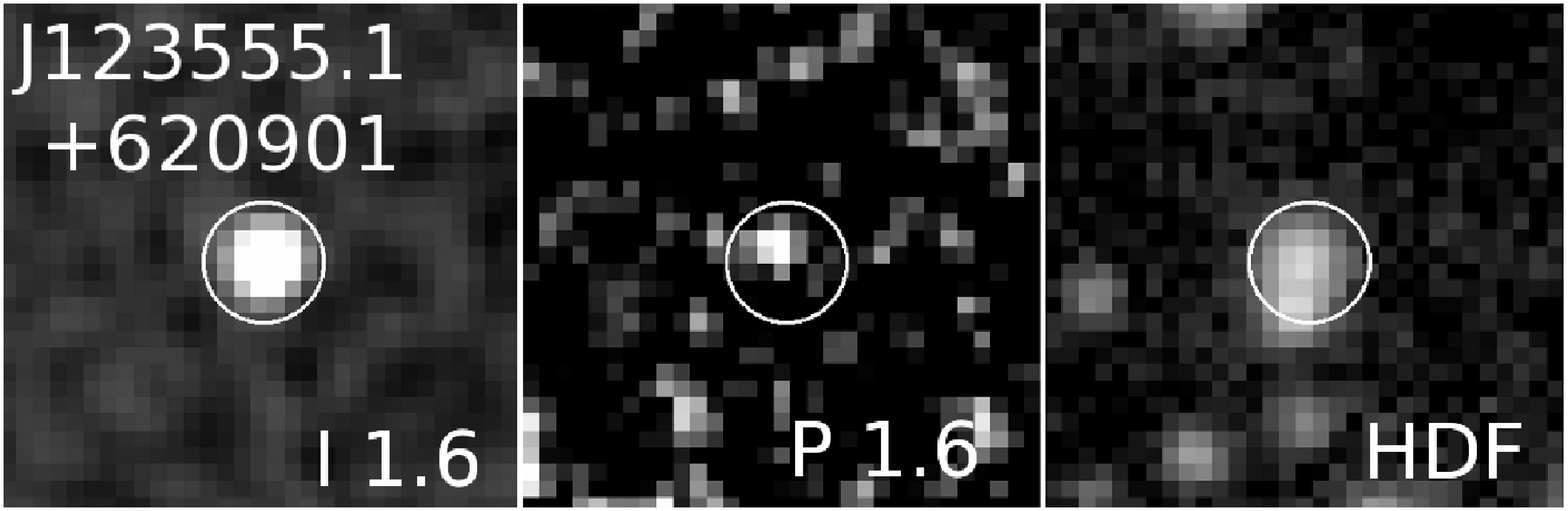}
\vskip .1in
\includegraphics[width=6.0cm]{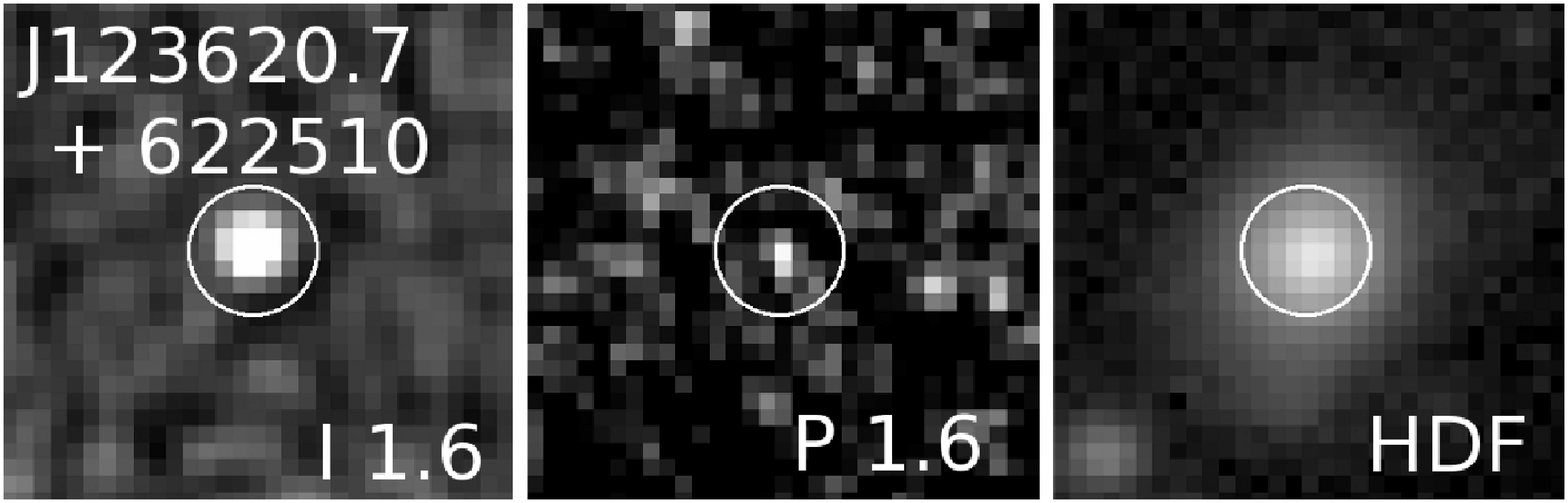}
\vskip .1in
\includegraphics[width=6.0cm]{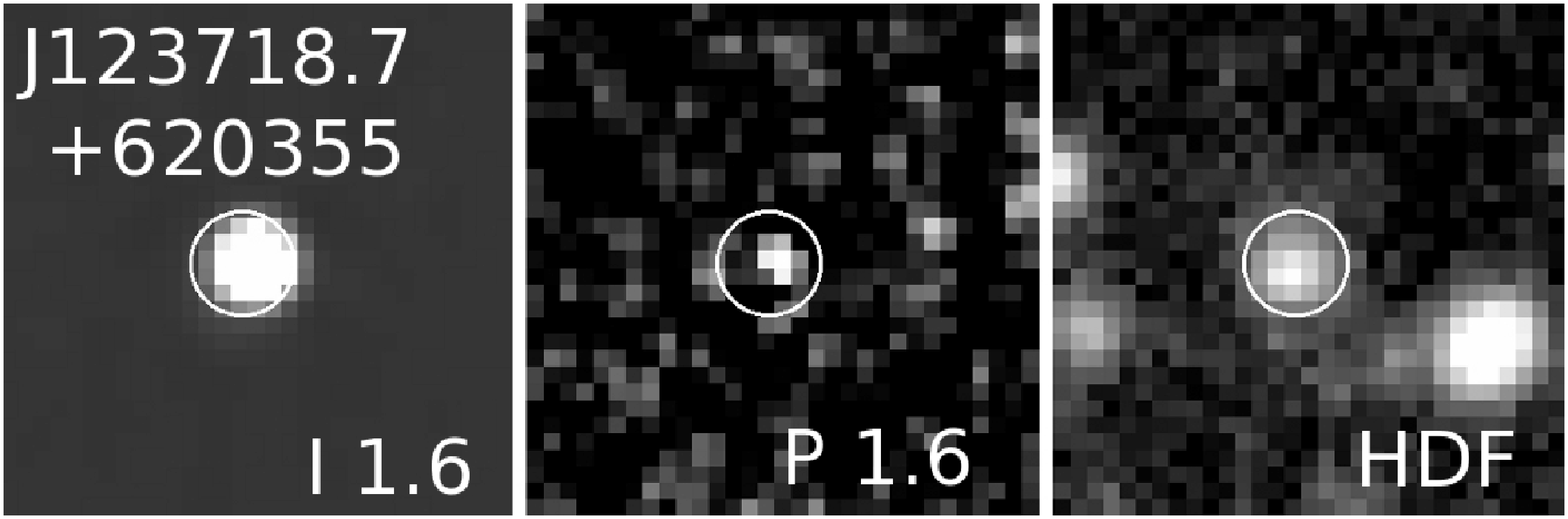}

\caption{\footnotesize{Compact sources. Field heights are all 15\arcsec. }}
\label{xtd1}
\end{center}
\end{figure}

\begin{figure}[!ht]
\begin{center}
\includegraphics[width=6cm]{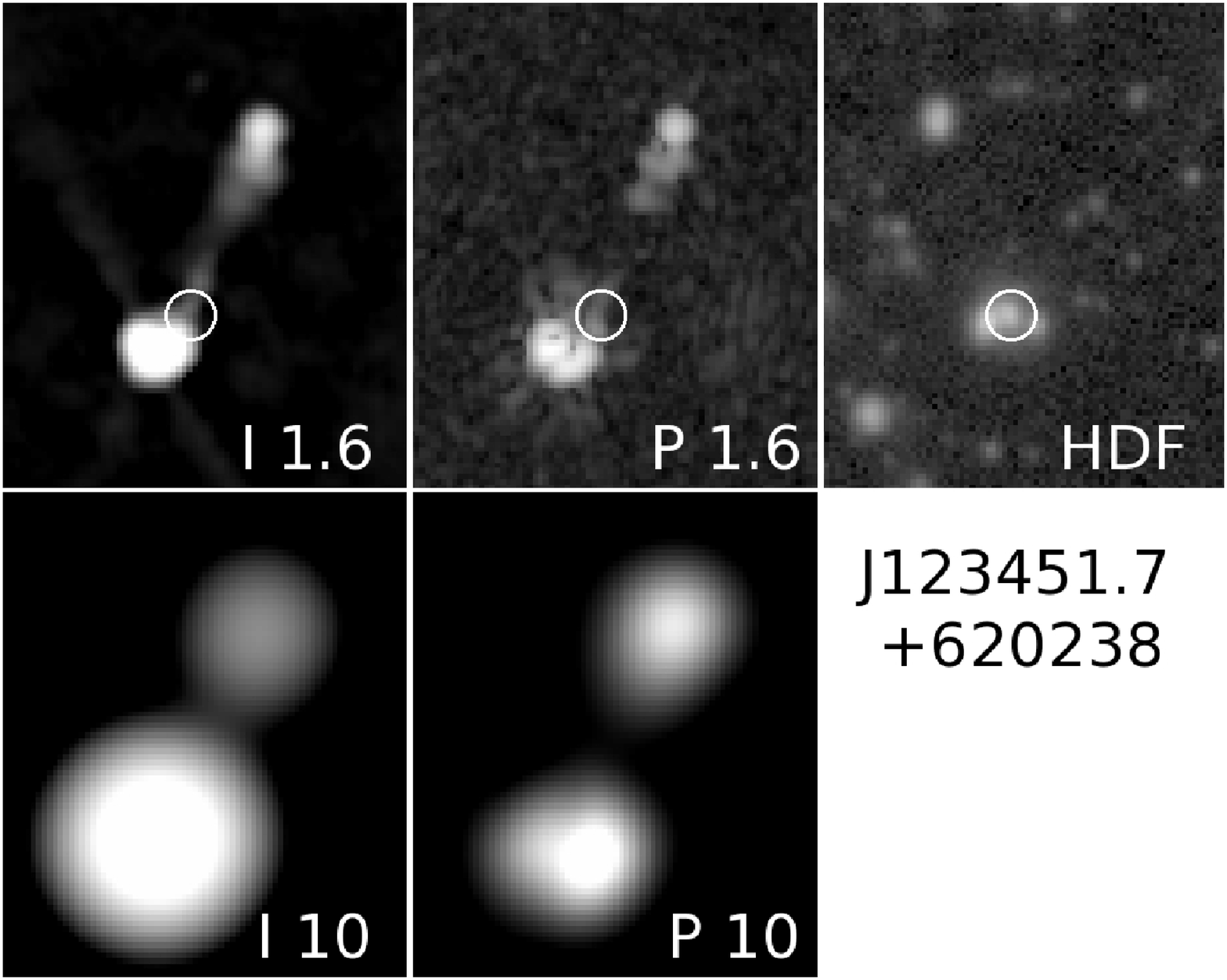}
\vskip .25in
\includegraphics[width=6cm]{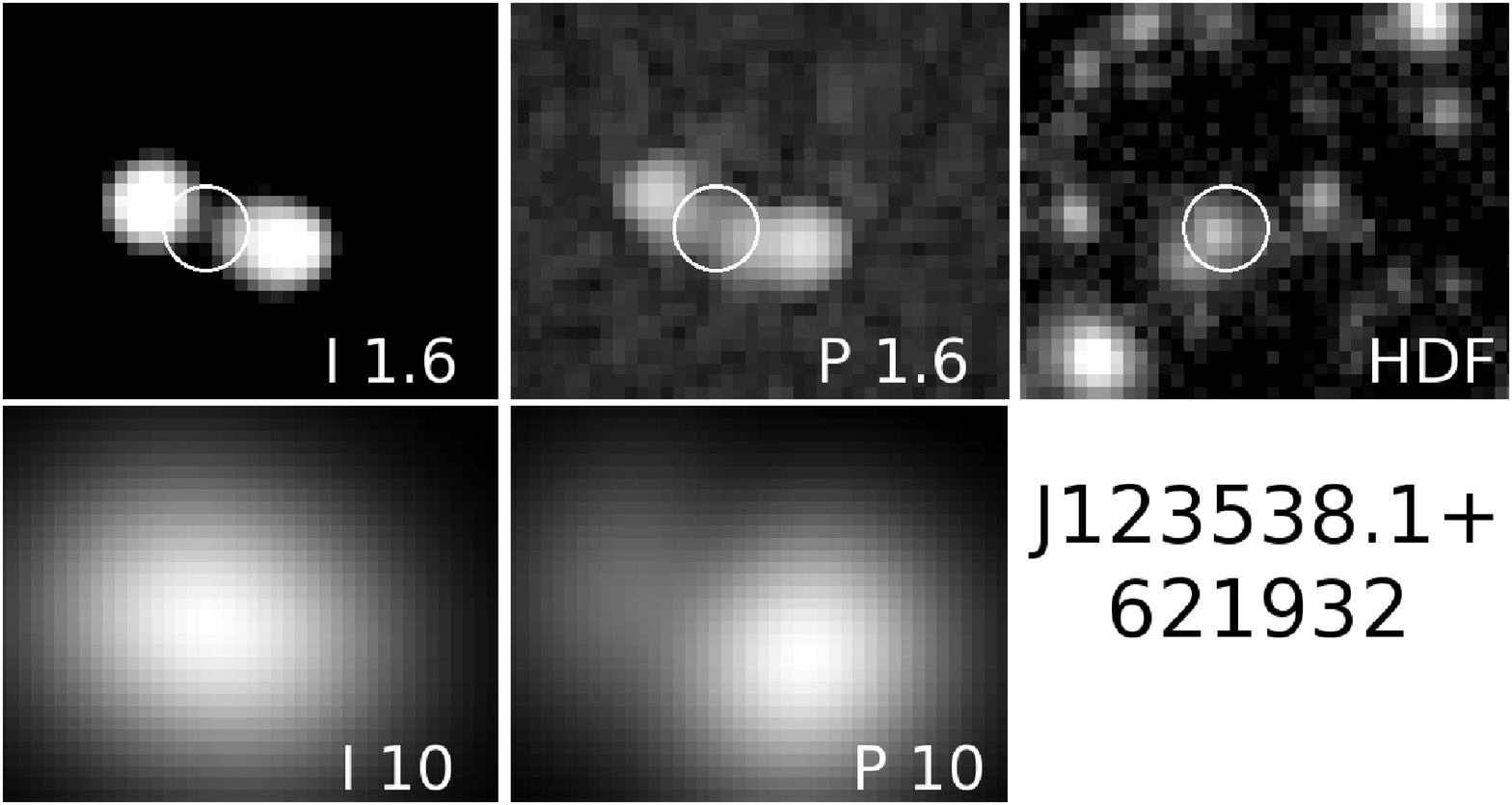}
\end{center}
\caption{\footnotesize{Extended sources. Field heights: top 45\arcsec, bottom 62\arcsec.}}
\label{xtd2}
\end{figure}

\begin{figure}[!ht]
\begin{center}
\includegraphics[width=6cm]{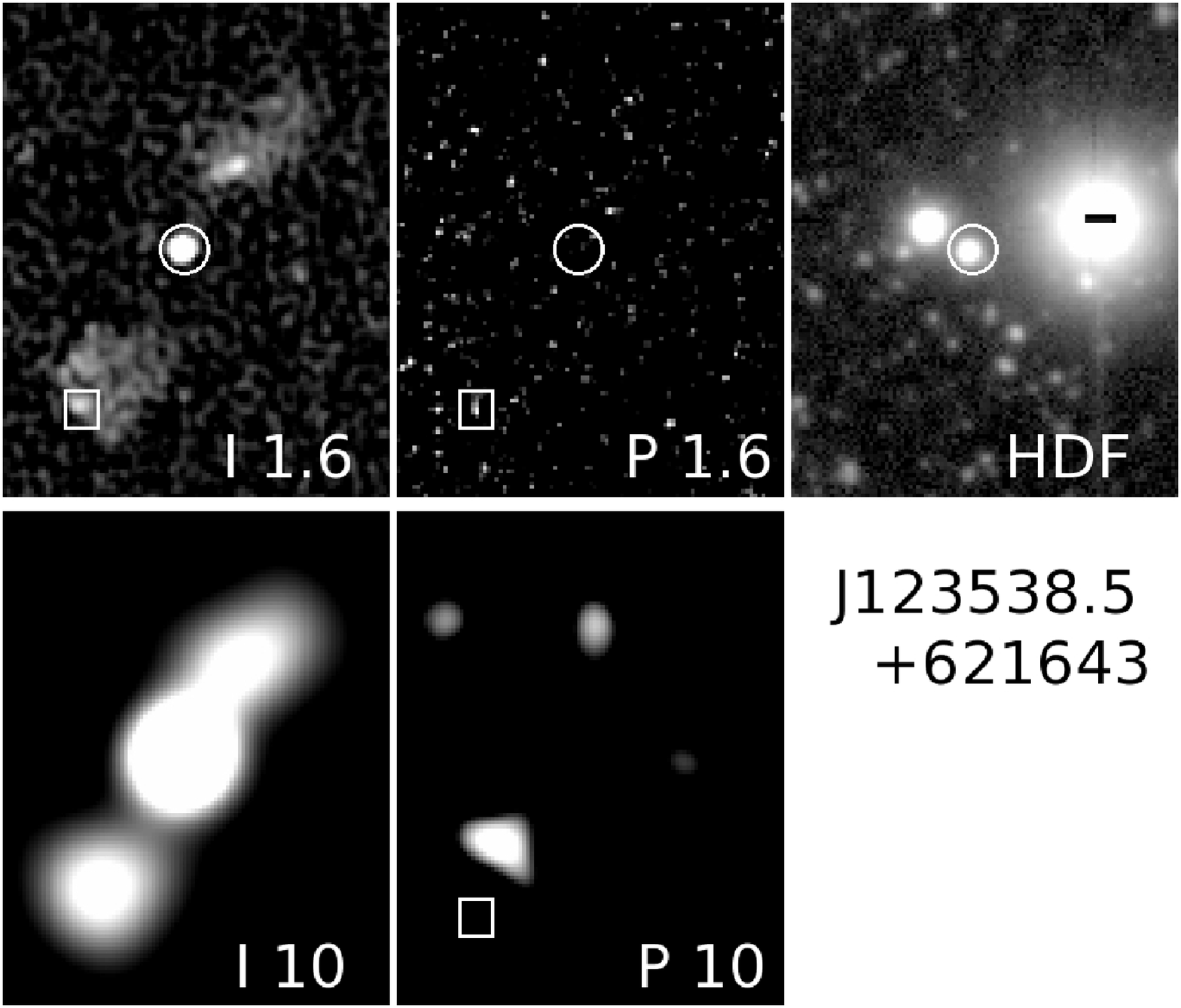}
\vskip .15in
\includegraphics[width=6cm]{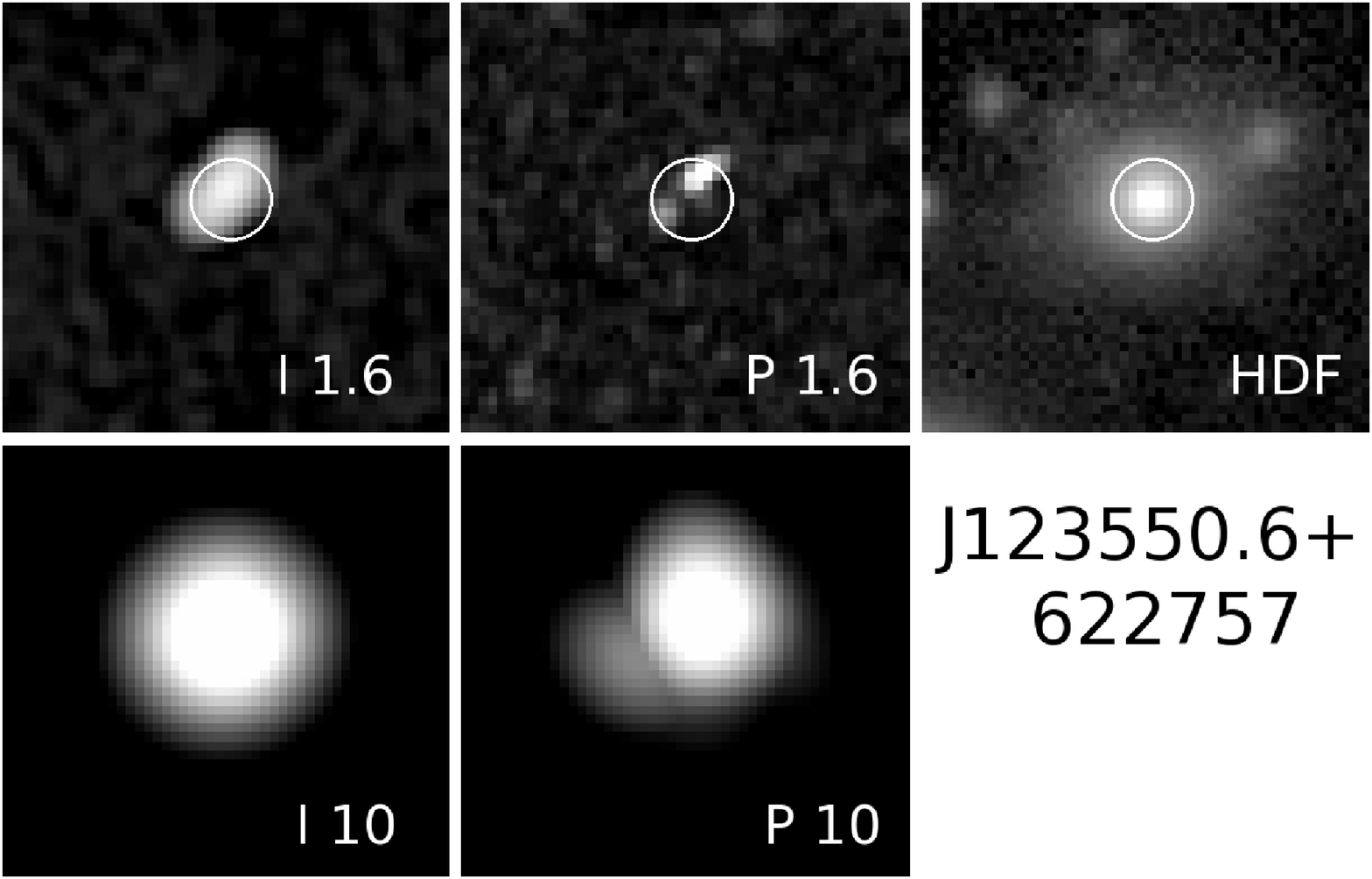}
\end{center}
\caption{\footnotesize{Extended sources. Field heights: top 61\arcsec, bottom 24\arcsec.}}
\label{xtd2a}
\end{figure}

\begin{figure}[!ht]
\begin{center}
\includegraphics[width=6cm]{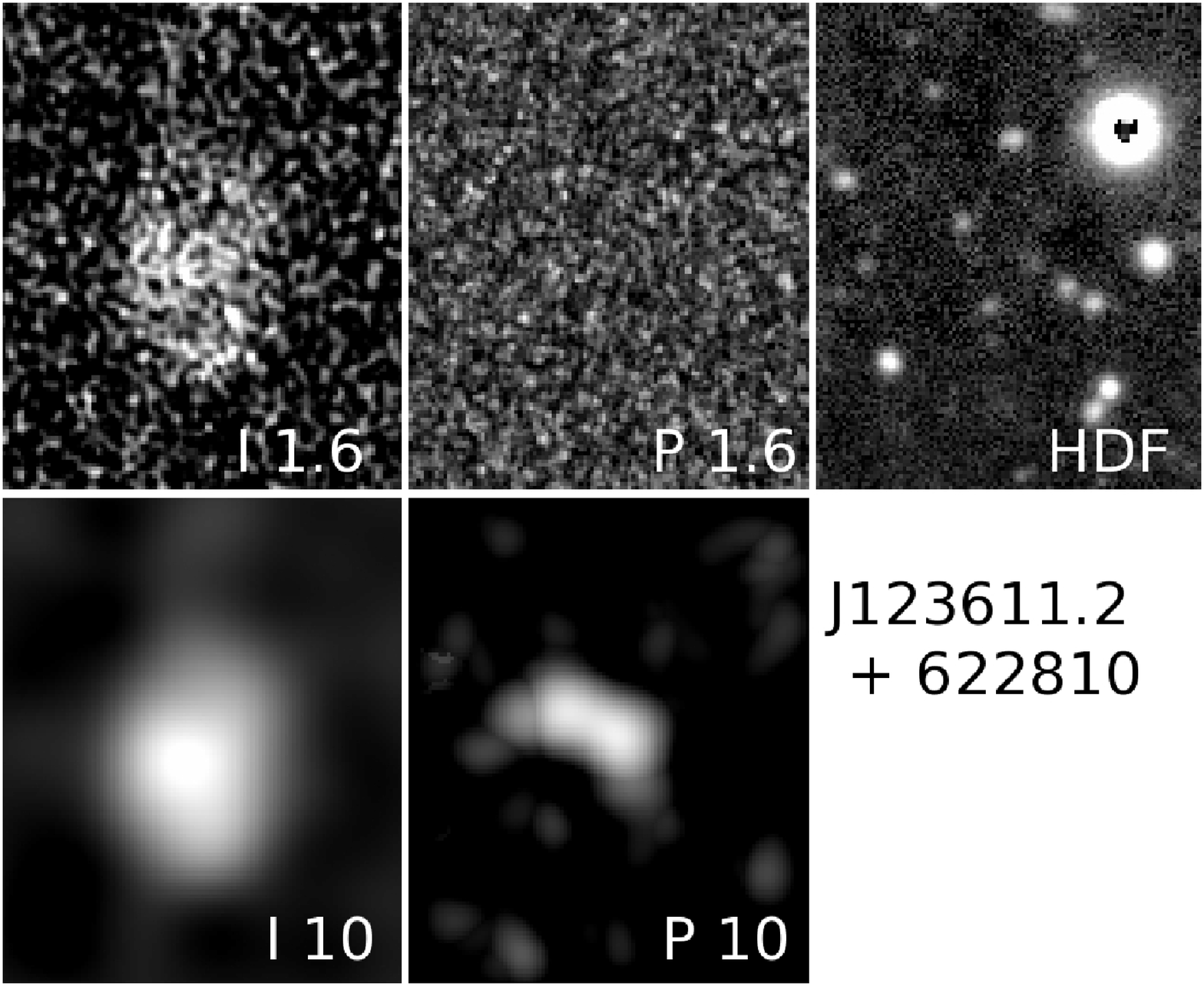}
\vskip .2in
\includegraphics[width=4.5cm]{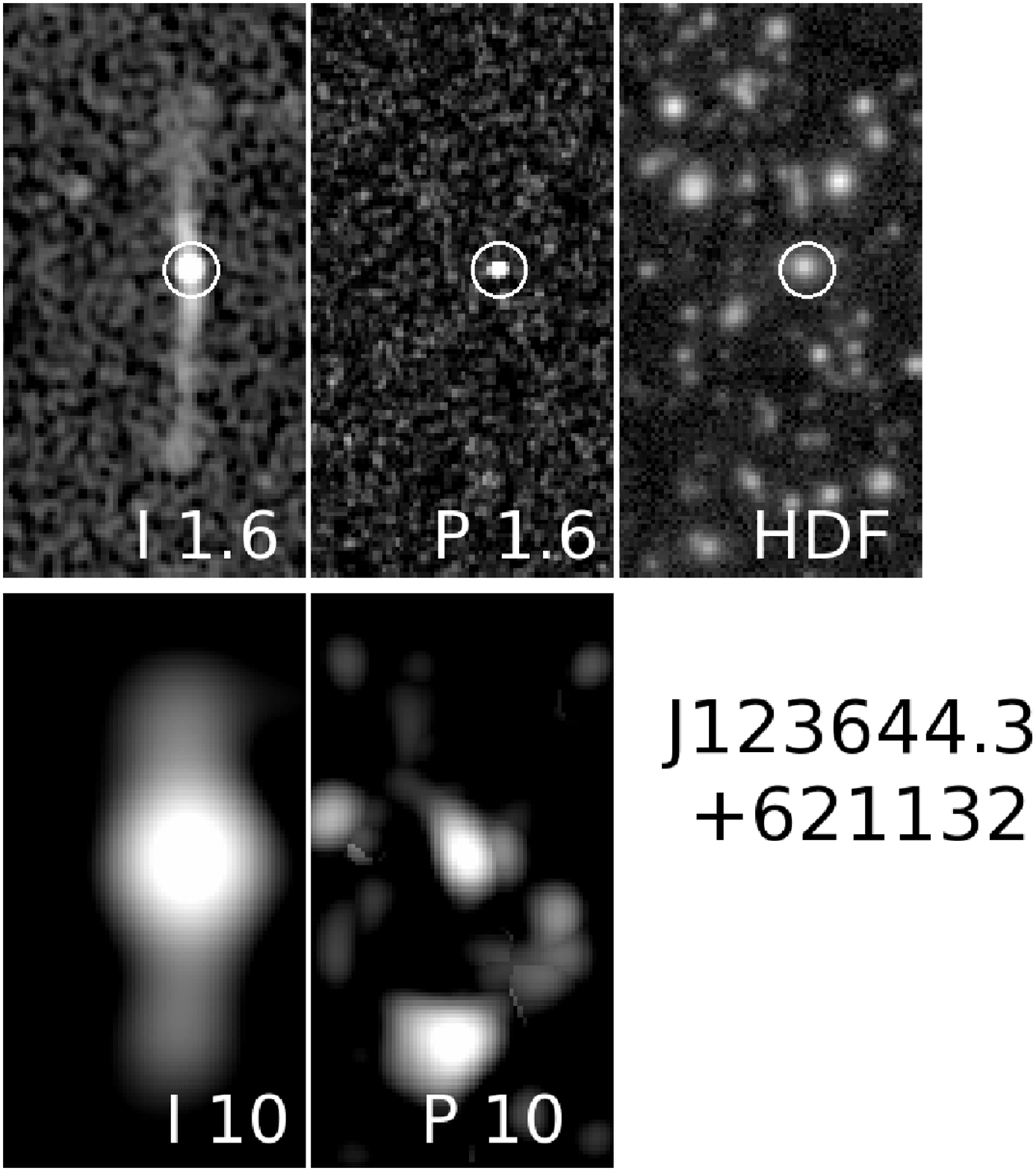}
\end{center}
\caption{\footnotesize{Extended sources, continued. Field heights: top 63\arcsec, bottom 64\arcsec.}}
\label{xtd3}
\end{figure}

\begin{figure}[!ht]
\begin{center}
\includegraphics[width=6cm]{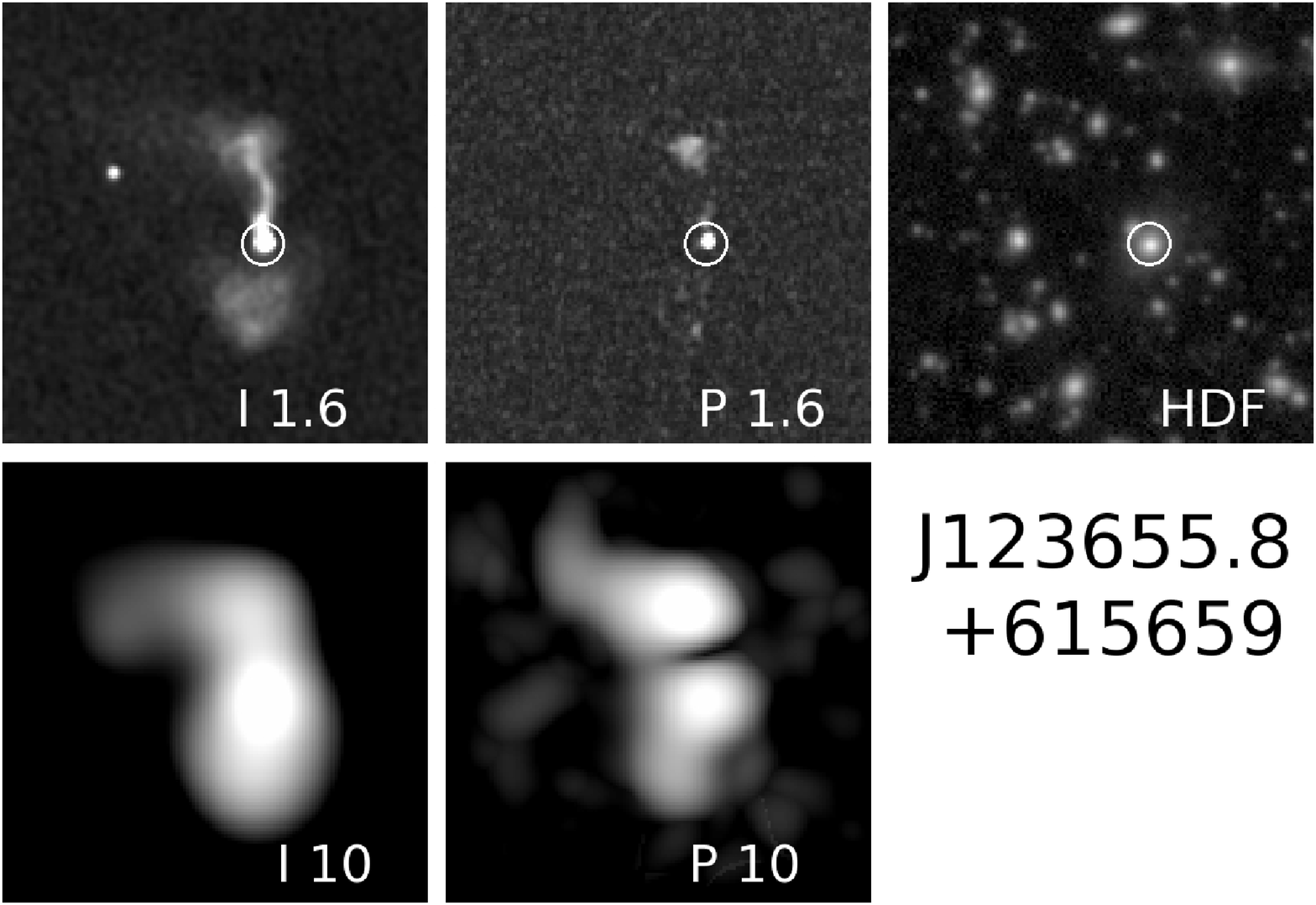}
\vskip .2in
\includegraphics[width=6cm]{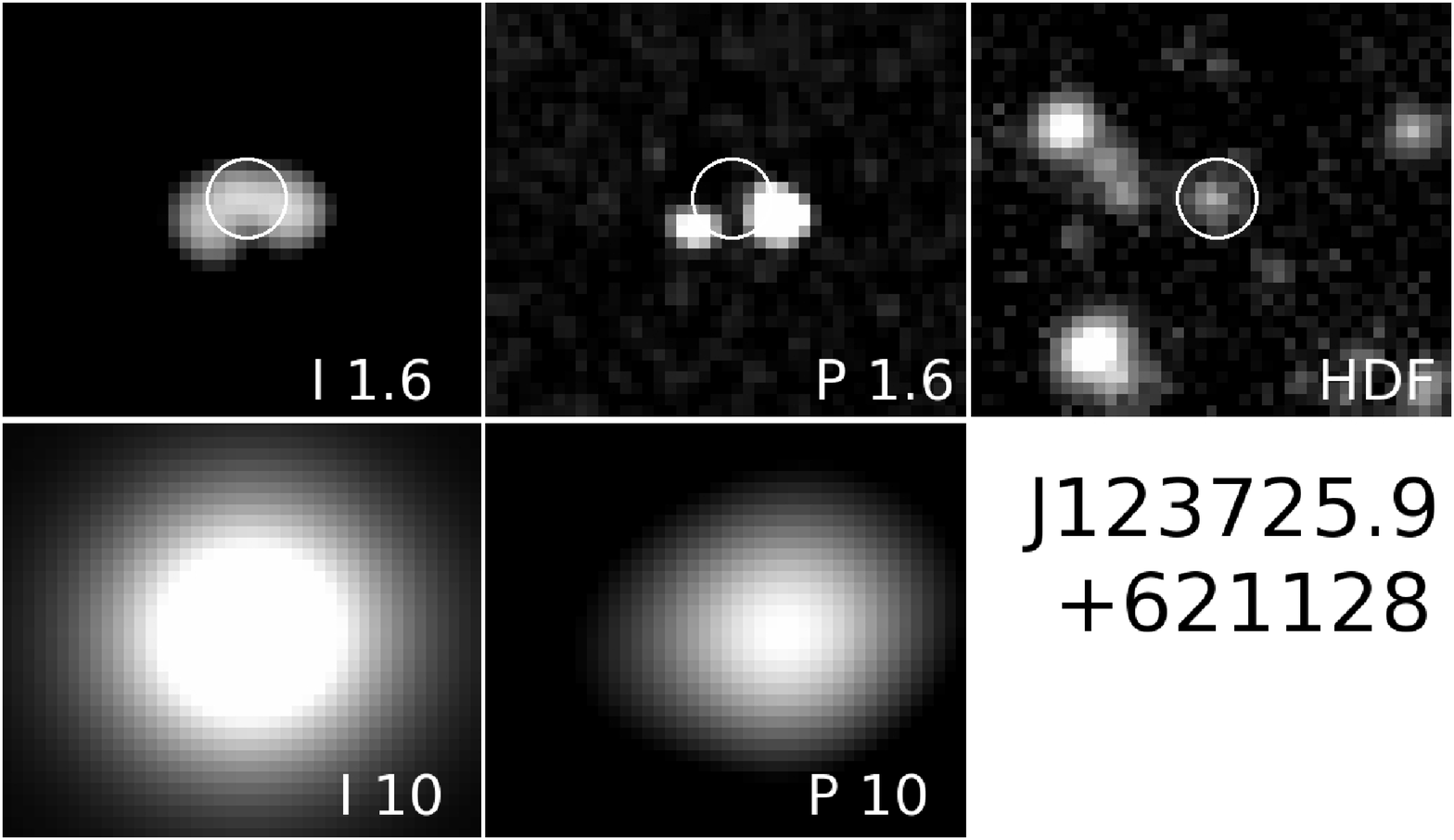}
\end{center}
\caption{\footnotesize{Extended sources, continued. Field heights: top 68\arcsec, bottom 19\arcsec.}}
\label{xtd3a}
\end{figure}

\begin{figure}[!ht]
\begin{center}
\includegraphics[width=6cm]{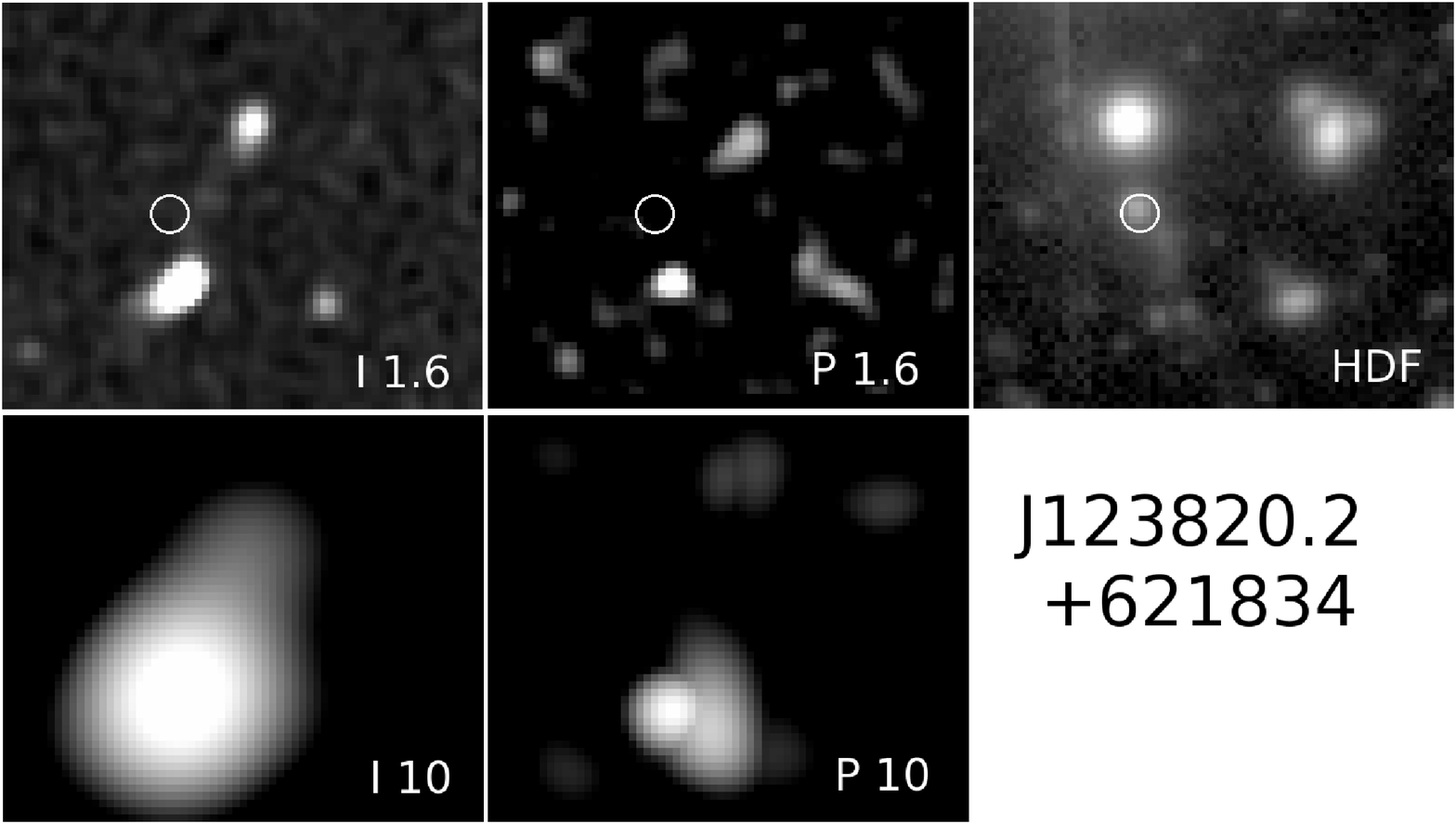}
\vskip .2in
\includegraphics[width=4cm]{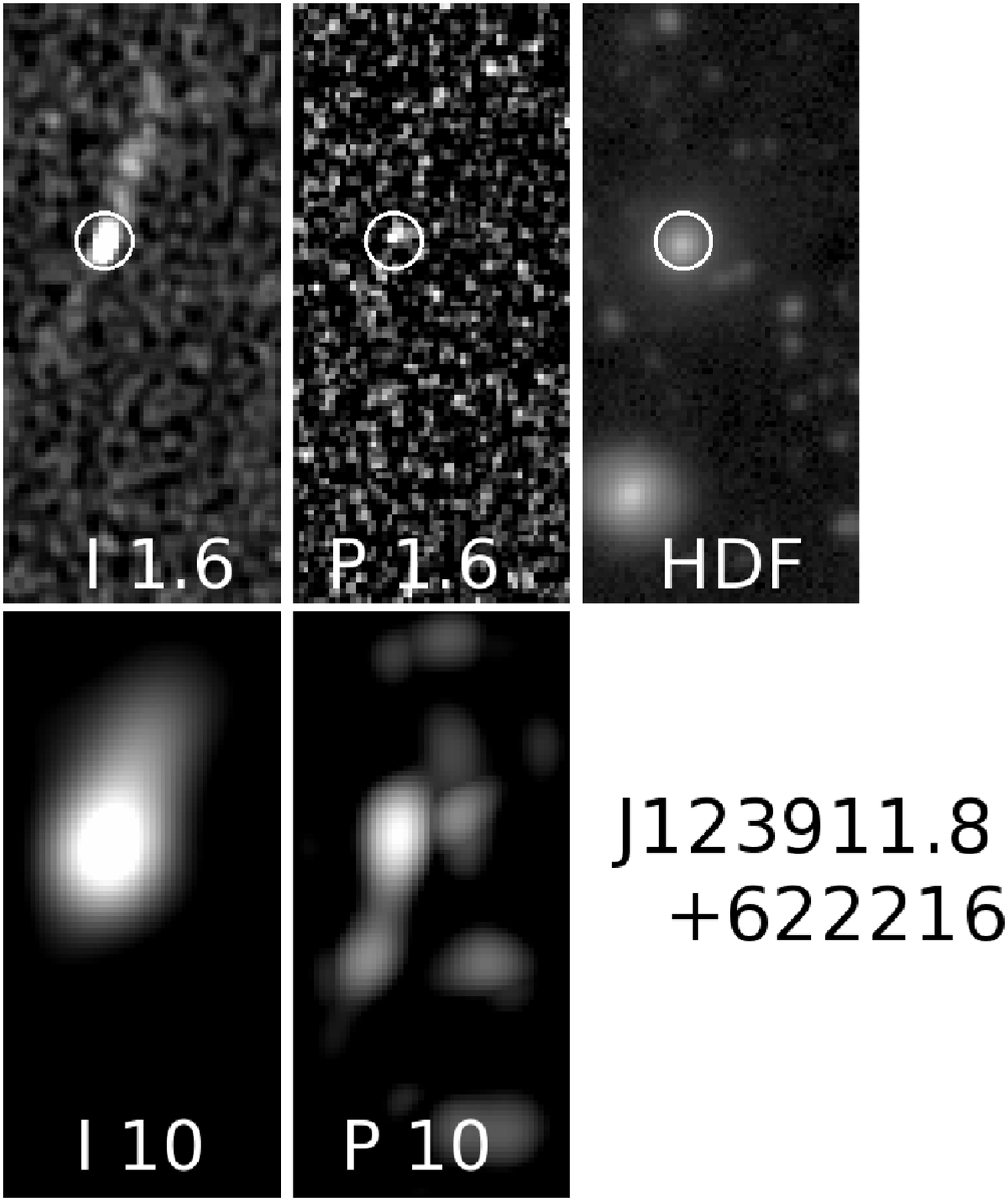}
\end{center}
\caption{\footnotesize{Extended sources, continued 2. J123820.2+621834 ID is tentative, since it is slightly off the radio major axis. Field heights: top 27\arcsec, bottom 57\arcsec.}}
\label{xtd4}
\end{figure}

Images of each detected source, along with the corresponding optical field, are shown in Figures \ref{xtd1}-\ref{xtd4}.  For extended sources, we show the images at both 1.6\arcsec~ and 10\arcsec~ resolution.



\section{POLARIZED SOURCE DISTRIBUTIONS}

\subsection{Number Counts}

Estimates of polarized number counts were done using the  1.6\arcsec~ resolution peak fluxes.  The advantages and disadvantages of this choice are discussed in Section \ref{resolution}. The reader should note that some discrepancies between these results and number counts at lower resolution are expected;  however, there is currently insufficient information to model the resolution dependence appropriately.   We also show the number counts both with and without including sources with high values of $|RM|$, as discussed in Section \ref{highRM}.

Because of the small statistics  (13 full resolution detections)  we calculated the 1.6\arcsec~ number count distribution in two complementary ways.  The first, ``Direct'' calculation looks at each source individually, and calculates the solid angle over which that source could have been detected, yielding a surface density contribution for this source.  The number counts are then derived by summing the surface densities from all detected sources with peak polarized fluxes greater than each limiting flux, P$_{limit}$.  The second method used a Monte Carlo approach, adopting a power-law form for the counts, and  minimizing the residuals from a test statistic derived from the data.  The two approaches give a generally consistent picture of the polarized number count distribution, and are described in detail below.

\subsubsection{Direct}  In this method, each source is handled individually, with the results combined at the end.  For each source, we calculated the primary-beam-corrected polarized flux P$^{1.6}_{pb}$, as listed in  Table 1.  We then calculate the area over which a source of that strength could have been detected, by finding the maximum distance from the field center at which its flux,{\em as reduced by the primary beam},  would be greater than our threshold of 14.5~$\mu$Jy. The maximum available area was based on our square image, 33.3\arcmin~ on a side, including a reduction in the available area because of  some small pieces of the image that were unusable due to sidelobe structures from strong sources. Finally, a reduction was made in the detectable area due to completeness, as shown in Figure \ref{models}. The maximum area over which an arbitrarily strong source could have been detected was 989 square arcminutes. 

\begin{figure*}[!ht]
\begin{center}
\includegraphics[width=10cm]{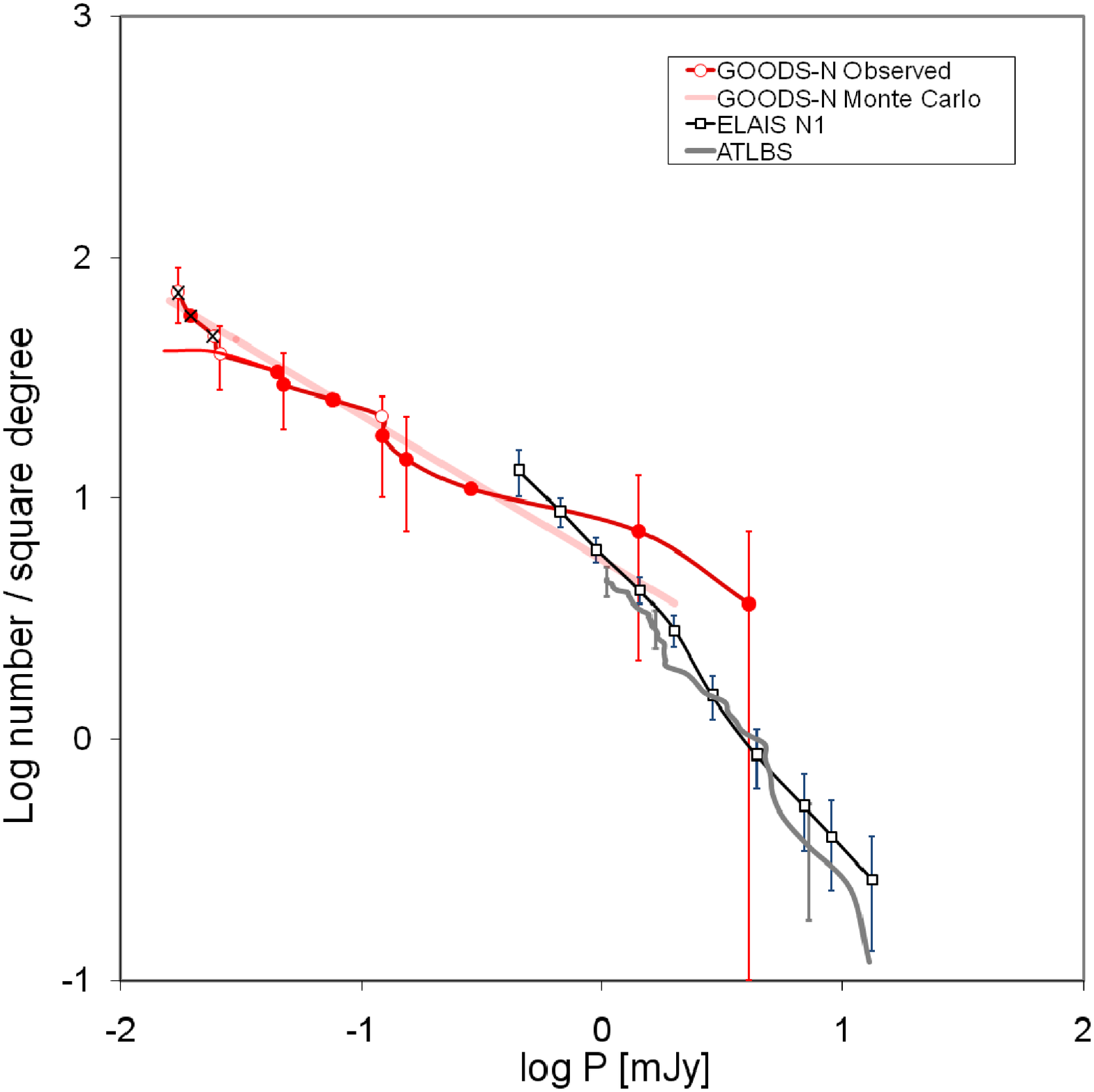}
\end{center}
\label{thecounts}
\caption{\footnotesize{Cumulative counts of polarized sources in GOODS-N field at 1.6\arcsec~ resolution, in red, with errors as described in the text.  Compact sources have open circles; filled circles represent extended sources, plotted at their peak polarized intensity at 1.6\arcsec~ resolution. Sources with Xs had high values of $|RM|$ as discussed in Section \ref{highRM}, and the dark red lines show the derived counts with and without these sources.   The thick pink line is the Monte Carlo fit to the GOODS-N data including all sources detected at 1.6\arcsec~ resolution.  Counts from 
surveys of the ELAIS-N1 and ATLBS fields at higher fluxes are also shown.  }
}
\end{figure*}

The resulting number counts are shown in Figure 10, along with earlier survey  data at higher fluxes from the literature.  These will be discussed in more detail in Section \ref{comparisons}.  As an estimate of the errors in the number counts, we assumed that each single source could either be missing, or that an additional source of that flux and position could have been present. Cumulative errors are then calculated by adding the surface density uncertainty contributions in quadrature.  Another number count distribution was also derived eliminating the sources with $|$RM$|>$150~\rpm, as a more conservative indicator of the number of sources that would be useful for foreground studies. This issue is discussed further in Section \ref{highRM}.


\subsubsection{Monte Carlo}
\label{montec}
 As an independent method of determining the number count distribution, we conducted a series of Monte Carlo trials assuming a power law distribution $$log~N(>P)~=~log(N_0)~+~\alpha\times~log(P/P_0),$$ where P$_0$ = 30$\mu$Jy. The trials were made by first setting up 20 logarithmically spaced polarized flux bins in the interval 10$\mu$Jy - 100mJy and then calculating, based on $N_0$ and $\alpha$, how many sources in each bin would be found in 100 square degrees.  The proper number of sources for each polarized flux bin were then  placed in random positions in the 100 square degree area, but counted only if they fell into a pre-designated area equal to that of the GOODS-N field, as described above.  Then, the polarized flux for each possible detection was degraded by the primary beam, based on its location in the virtual GOODS-N field, and kept as a possible detection only if the degraded flux was above the detection threshold of 14.5$\mu$Jy.  Finally, only a fraction of the possible detections were actually counted, based on the completeness at that polarized flux level, as shown in Figure \ref{models}. This whole procedure was then repeated 10,000 times for each value of $N_0$ and $\alpha$.

To compare the likelihood of different assumed values for N$_0$ and $\alpha$, we constructed a figure of merit using the number of Monte Carlo sources detected above and below P~=~100~$\mu$Jy  (N$_+$ and N$_-$, respectively), where the actual number of observed sources were 5 and  7, respectively. Two of the 14 sources listed in Table 1 are not counted here;  J123611.2 is not detected at full resolution, and to be slightly conservative, we eliminated J123911.8 as potentially spurious because of its very high fractional polarization.
We then defined $\chi^2 \equiv (N_- - 7)^2/7 + (N_+- 5)^2/5$ and counted the number of trials, out of 10,000, where $\chi^2 < 1$ (expected 40\% of the time for two degrees of freedom). The maximum likelihood combination of parameters was
log(N$>$P)~=~log(45/square degree)-0.6* log(P/30$\mu$Jy), and is plotted in Figure 10. The number of ``successful'' trials was $\sim$4700 for the best fit, when 4000 were expected if the model fit the data.

The Monte Carlo test was repeated, eliminating the three sources with high $|RM|$ values.  In order to modestly improve the power of the statistics, we shifted the cutoff P to 150~$\mu$Jy, for which N$_+$ and N$_-$ are then 4 and 5, respectively. The best fit was for  log(N$>$P)~=~log(31/square degree)-0.63* log(P/30$\mu$Jy) and the number of trials with $\chi^2<$1 was ~$\sim$4300, with an expected 4000.

Again, we note that these apply to source counts at 1.6\arcsec~ resolution, using peak fluxes.

\subsubsection{Comparison to polarized source counts in the literature}
\label{comparisons}
In Figure 10,
we compare our cumulative polarized number counts to two shallower, lower resolution surveys at 1.4~GHz. As noted earlier, there are differences expected in the counts at different angular resolutions, although it is not possible to model these at this time.  At the simplest level, the polarized fluxes for our detected sources would  go up when integrated over the entire source.  If nothing else changes in terms of detectability, then the above values for N$_0$ normalized back to 30$\mu$Jy would be N$_0$=68(48)/square degree, with (and without) the high $|$RM$|$ sources.

 The first comparison survey is that of the ELAIS N1 field  \citep{elais1,elais2}, covering an area of 15.16 square degrees, based on observations at the Dominion Radio Astrophysical Observatory.  The noise in Stokes Q, U was 78~$\mu$Jy/beam, at a resolution of $\approx$50\arcsec$\times$60\arcsec (N-S).  Below P$\sim$500~$\mu$Jy their completeness correction is larger than a factor of 10 \citep{grantphd},  and the count estimate becomes very uncertain, so the results are presented here only above that 500~$\mu$Jy level. The second survey is of the ATLBS field \citep{atlbs} covering an area of 8.42 square degrees, based on observations at the Australia Telescope Compact Array, with a noise of $\sim$85$\mu$Jy/beam at a resolution of 50\arcsec.  The claimed detections of polarized sources appears to be problematic at low fluxes because of the detection threshold of 1$\times\sigma$;   this threshold should be exceeded 60\%  of the time, in the presence of noise alone.  We therefore adopted a more conservative detection limit of p=1~mJy from the ATLBS survey to reconstruct the number counts from the data in their Table 2.   No correction was then necessary for incompleteness due to the primary beam, because it is not important given our higher assumed detection threshold.  A more detailed analysis of the problems with the ATLBS counts is presented in \cite{stack}. 

The much larger survey areas in the ATLBS and ELAIS N1 fields allow them to get significantly better statistics on the relatively rare sources with p$>$1~mJy.  Our results are consistent with both of these polarized source counts, albeit with quite large errors.  However, where our statistics begin to improve, at p$<$1~mJy, it is clear that the steep slope of the number counts in the ATLBS and ELAIS N1 fields does not continue.  There is an apparent break in the cumulative count slope, from $\alpha\sim$-1.4 above p=1~mJy to $\alpha\sim$-0.6 below 1~mJy, as seen in Figure 10.

\subsubsection{Comparison to polarized source count models}

\begin{figure*}[!ht]
\begin{center}
\includegraphics[width=9.3cm,height=9cm]{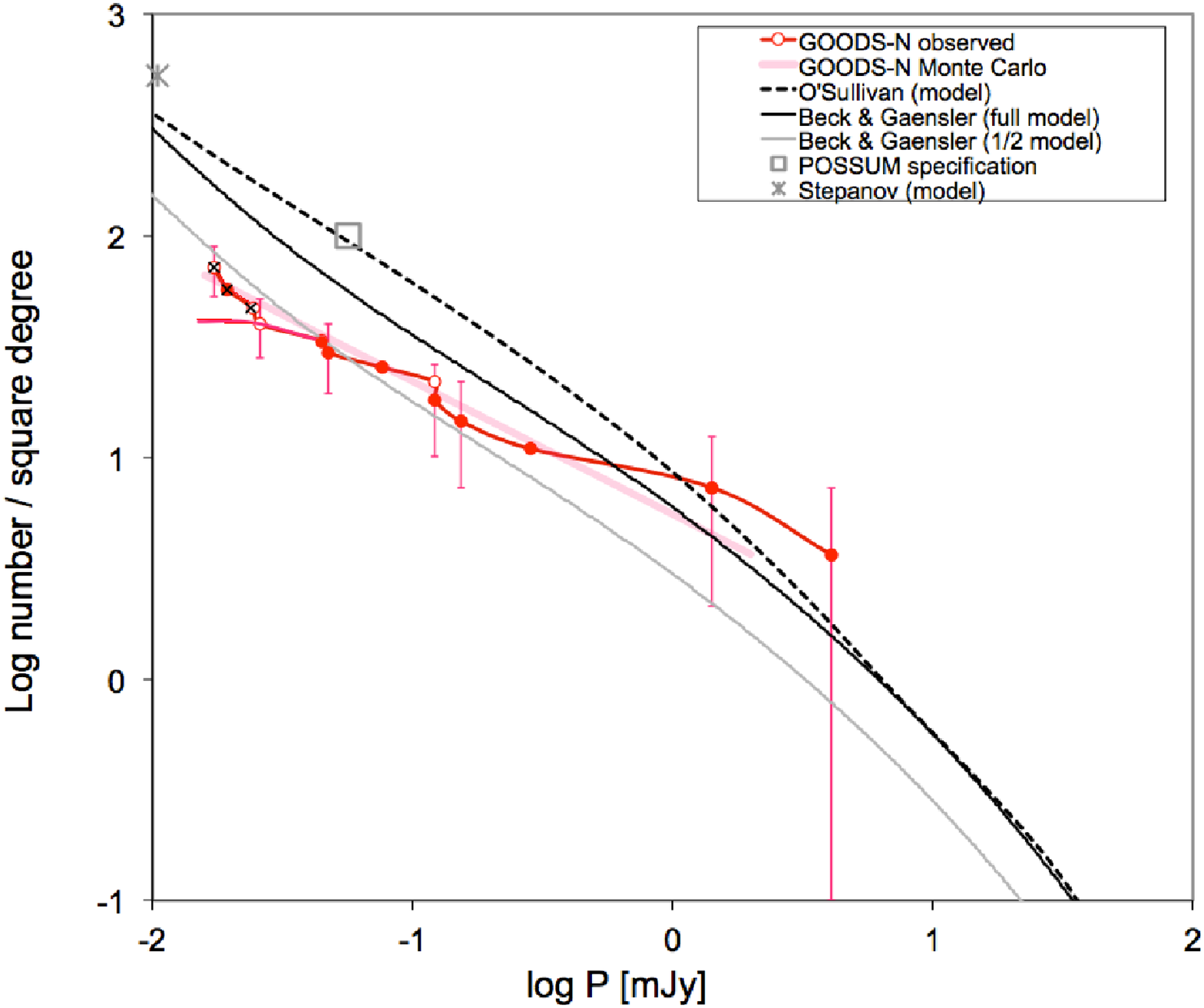}
\end{center}
\caption{\footnotesize{Comparison of cumulative GOODS-N counts at 1.6\arcsec~ resolution with models in the literature. Symbols and lines defined as in Figure 10. See text for further description.}}
\label{lowRMcounts}
\end{figure*}

We now compare the newly derived GOODS-N counts with several predictions from the literature.  \cite{beckg} proposed an all-sky survey of Faraday rotation for the upcoming Square Kilometer Array (SKA), and produced number count projections for these sources.  They used the total intensity counts of \cite{hopkins} and \cite{seymour}, and a polarized fraction distribution derived from I$>$80~mJy sources from the NVSS catalog \citep{NVSS}, which they modeled as two gaussians centered around a fractional polarization of $\sim$3\%. Two curves from this work are shown in Figure \ref{lowRMcounts}.
 The upper, black curve is the total predicted number of polarized sources, using an analytic expression kindly provided by B. Gaensler. The lower grey curve is a factor of two lower, as published in \cite{beckg}, and accounts for their estimate of those sources for which RMs can be reliably determined.  From our current GOODS-N data, all of the RMs above p~$\sim$~30~$\mu$Jy would be suitable for foreground Faraday studies, so the upper curve provides the most appropriate comparison.  At $\sim$30~$\mu$Jy the prediction is a factor of 2-3 higher than the GOODS-N observations;  the Beck \& Gaensler lower (1/2 count) curve is consistent with the data, but for reasons that don't apply here.  Below $\sim$30~$\mu$Jy, the situation is less clear.  If the high $|$RM$|$ sources are not included, the number counts may become even flatter, reducing them further below the predicted values.   

\cite{osullivan} also derived a model for the faint polarized source population incorporating separate fractional polarization distributions for FRI and FRII sources, and a luminosity dependent mean fraction.  Using their formulation, they found good agreement with the polarizations from the NVSS, although they could not account for the flattening of the differential counts (more polarized sources than expected) reported at the lowest fluxes in the ELAIS-N1 survey.  However, as we note above, the ELAIS-N1 survey did not include errors associated with the uncertainty in the completeness correction, so it is not clear whether there is an actual discrepancy with \cite{osullivan} even down to the p=$\sim$300~$\mu$Jy level.  Comparison of their model with the GOODS-N counts in Figure \ref{lowRMcounts}, however, shows that the predictions strongly overestimate the number counts below 300~$\mu$Jy.

A different approach, working directly with observed polarized number counts, was utilized by \cite{step08}. They extrapolate the \cite{elais1} results down to 10~$\mu$Jy, for an estimate of $\sim$530 polarized sources per square degree suitable for foreground RM studies.  This is a factor of $\sim$10-20 above the counts expected from extrapolating the GOODS-N observations presented here down to 10$\mu$Jy.

Finally, we compare the observed density of sources at p$\sim$50~$\mu$Jy to that expected in the specifications for the Polarization Sky Survey of the Universe's Magnetism (POSSUM).~\footnote{POSSUM is planned to be carried out commensally with the Evolutionary Map of the Universe (EMU) Survey \citep{norris} at the Australian SKA Pathfinder (ASKAP).}  At the 5$\sigma$ level of 50~$\mu$Jy, POSSUM expected $\sim$100 sources per square degree \citep{possum} with RMs suitable for foreground screen experiments;  the GOODS-N 1.6\arcsec~ counts suggest that 
$\sim35\pm$10 per square degree is a better estimate, using our 1.6\arcsec~ counts directly.  Because POSSUM will have a 10\arcsec~ beam, however, the counts might be enhanced by the additional polarized flux, and by picking up additional extended sources  such as J123611.2+622810 and J123538.5+621643, which  are both detectable and meet the low $|$RM$|$ criterion only at 10\arcsec~ resolution.  At the same time, the counts would go down  by 
$\sim$35\% if POSSUM adopts a more conservative 10$\sigma$ detection limit, in order to search over a very broad range in Faraday space.

\subsubsection{Elimination of high $|RM|$ sources}
\label{highRM}

{ The three weakest polarized sources have anomalously high values of $|RM|$ compared to the others. Figure \ref{models} (right) shows that such high values are common below the detection threshold of 14.5$\mu$Jy, where spurious sources are present. Above the threshold, spurious detections are rare ($<$1\%) and the maximum $|$RM$|$ observed in our noise experiments was only 35~\rpm.  Thus, eliminating sources on the basis of high $|RM|$ alone could produce a scientific bias.  However, there are other reasons to be suspicious.  While heavily depolarized sources might show intrinsically high $|RM|$, that is not the case here.  The fractional polarizations are approximately 11\%, 25\% and 47\% .  These are considerably higher than the other fractional polarizations observed, which itself raises further doubts about the reality of these detections.  Finally, J123620.7+622510 and J123558.5+621643 appear marginal on Figures \ref{xtd1} and \ref{xtd4}, respectively.  Given this uncertainty, we have chosen to report the results as observed, and discuss the implications of either including or excluding the weak sources for both Faraday foreground  and source population studies.}

\begin{figure}[!ht]
\begin{center}
\includegraphics[width=7cm]{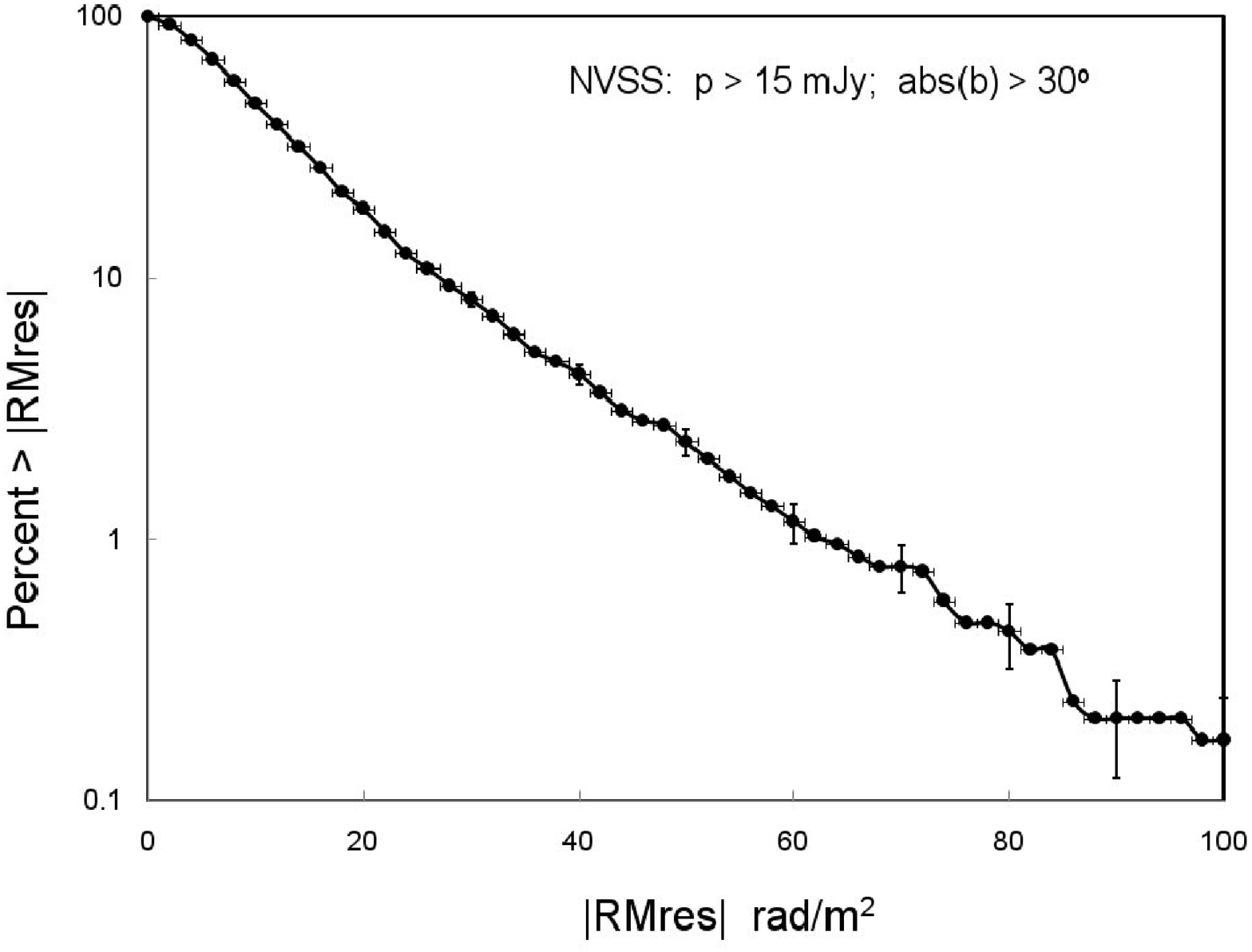} \hskip 0.25in
\includegraphics[width=7cm]{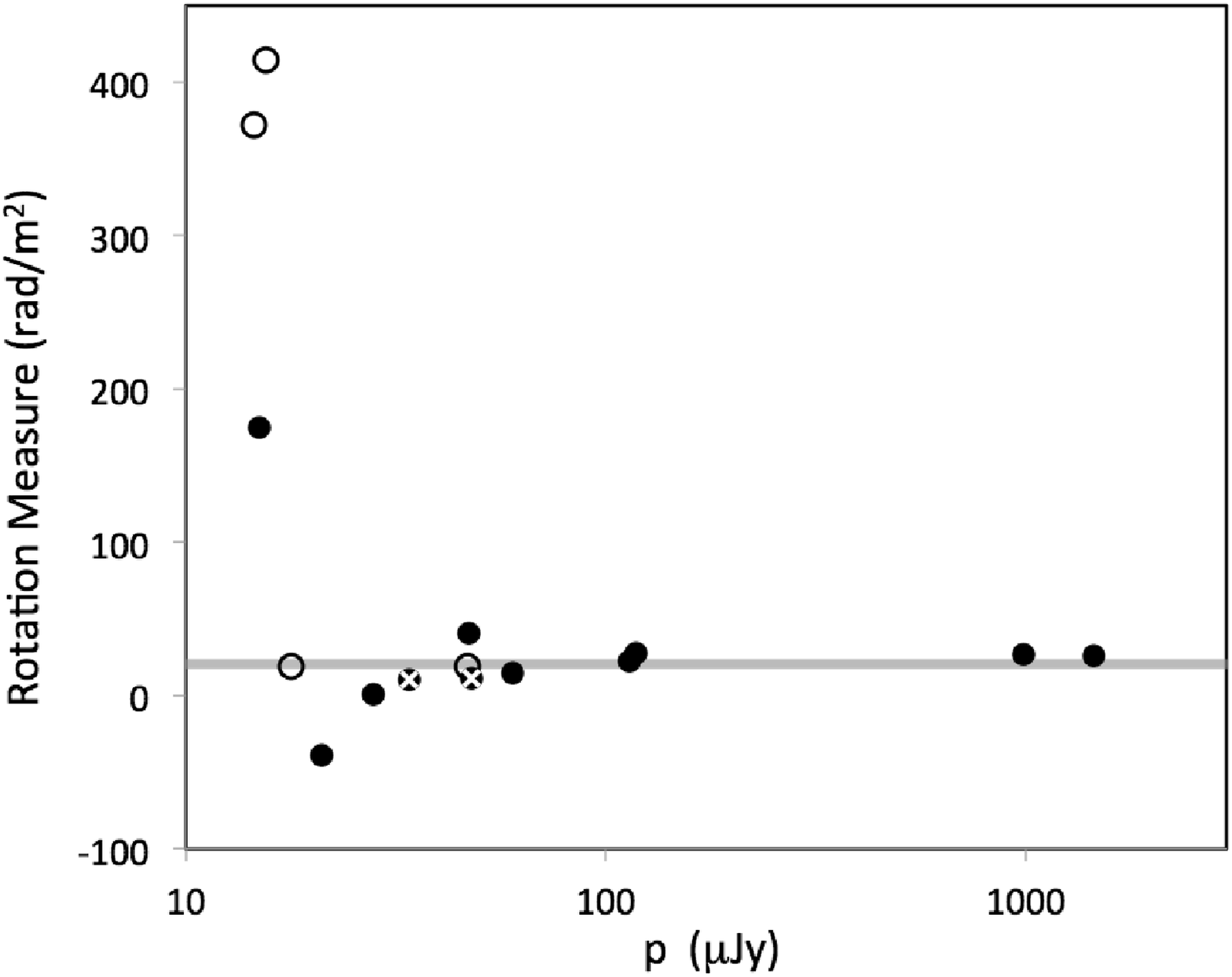}
\end{center}
\caption{\footnotesize{{\em Left:} Cumulative percentage of observed residual RMs of NVSS sources \citep{taylornvss} after subtraction of the mean RM in a 3$^{o}$ radius circle around each source. {\em Right:} RM as a function of observed p for GOODS-N.  Open circles are compact, filled circles are from extended sources, all sampled at 1.6\arcsec~ resolution.  The two filled circles with white X's are from 10\arcsec~ resolution images of J123611.2+622810, which is not detected at 1.6\arcsec, and of J125338.5+621643, which is also plotted at 1.6\arcsec resolution, where a different polarized component is seen. The line at 20 rad/m$^2$ indicates the median RM from the NVSS survey in a 3$^{o}$ radius circle centered on the GOODS-N field.}}
\label{RMnvss}
\end{figure}

One of the major goals of polarization surveys is to provide background sources for measuring foreground Faraday screens, so we determined how the number counts would change if we included only those sources suitable for such measurements. The key requirement is that the {\em intrinsic} value of the rotation measure  (RM), (i.e., local to the source), is sufficiently small in absolute value. Since it is not possible to tell, for each individual source, whether its observed RM is intrinsic or due to the foreground screen, we can look at the overall statistics of background RMs. As part of a separate study, we used the NVSS catalog of RMs \citep{taylornvss} and for each source where p$>$15~mJy and $|$b$|>$30$^o$, we calculated an {\em RM residual} by subtracting the median RM of all sources within a 3$^o$ radius circle around the source.  The results are shown in Figure \ref{RMnvss};  the median   $|$RM$_{residual}|$ is 9.2~\rpm~; only 1\% of background sources have 
 $|$RM$_{residual}|>$60~\rpm.  Thus, for strong enough sources at high galactic latitude, polarized source searches would recover 99\% of the sources by covering RMs within 60~\rpm~ of the average local RM value.

 Figure \ref{RMnvss} also shows the dependence of RM on p from the GOODS-N sample.  Above P=25$\mu$Jy, the mean and rms of the RMs are 20\rpm~ and 11\rpm, respectively.   The mean is the same as the  foreground galactic value of 20~\rpm, determined by taking the median of all RMs from the NVSS \citep{taylornvss} within a circle of radius 3$^{o}$ around the GOODS-N field center. We detect 3 sources out of 15\footnote{There are 15 sources in this reckoning because J125338.5+621643 is plotted twice, once at each resolution; it shows a residual RM of 155 \rpm at 1.6\arcsec, and -10 \rpm at 10\arcsec resolution.} showing large deviations in RM from the local Galactic foreground. From Figure 12, only  $\sim$0.1 such a high $|RM|$ source is expected at random.  These anomalous results are likely tied to the sources themselves { (either physical or instrumental)}, and not the foreground, because the high $|$RM$|$s come from  the three weakest sources in both polarized  and total intensity. 
 
 Whatever the cause of the high $|$RM$|$s, to the extent that these are intrinsic to the source,  they would not be useful for measuring foreground Faraday screens. Whether or not to eliminate such sources is a matter of scientific judgement.  For galactic studies, e.g., an isolated high $|$RM$|$  source might indicate a small intervening HII region, and should not be eliminated.  For other studies, such as the outskirts of galaxy clusters, including such sources would bias the Faraday modeling.  Other evidence for intrinsic, rather than foreground contributions to the RM need to be considered, such as distortions in the radio galaxy structure, spuriously high fractional polarizations, or polarized fluxes near the detection limit.

{For source population studies, including the three weakest polarized sources changes the results at the low end of our distribution.  Therefore, in Figures 10 and 11, we show the surface density of sources with and without the three weakest ones; not counting the three weakest sources results  in a drop by  a factor 1.75 at 15~$\mu$Jy..  We also ran the Monte Carlo simulations in Section \ref{montec} without the three weakest sources, resulting at a drop by a factor of 1.4 at 15$\mu$Jy.  The results we quote in this paper do include the three sources; this is a conservative approach for comparison with the literature, since our source densities are already lower than other estimates.  Our counts are also likely affected by resolution, as discussed below.  Readers should note  that the source densities could be even lower than we report, if the three weakest detections are spurious.}

\subsubsection{Resolution effects}
\label{resolution}
Number counts and fractional polarization distributions depend on the angular resolution of the survey.  At the simplest level, sources can be blended at low resolution, reducing the number of ``source'' detections in both total intensity and polarization.  Additional complications arise in polarization, where blending of regions with different polarization angles and/or RMs will result in depolarization, and the subsequent loss of sources that fall below the detection limit.  On the other hand,
low surface brightness polarized regions may become detectable at lower resolutions, raising the number counts.  It is important to recognize the existence of these issues when comparing the current GOODS-N results with other surveys from the literature, which typically have resolutions $\sim$50\arcsec.  However, without more detailed models of the polarization structure of the underlying source populations it is not possible to quantify these effects, nor even determine their sign.

\begin{figure}[!ht]
\begin{center}
\includegraphics[width=7cm]{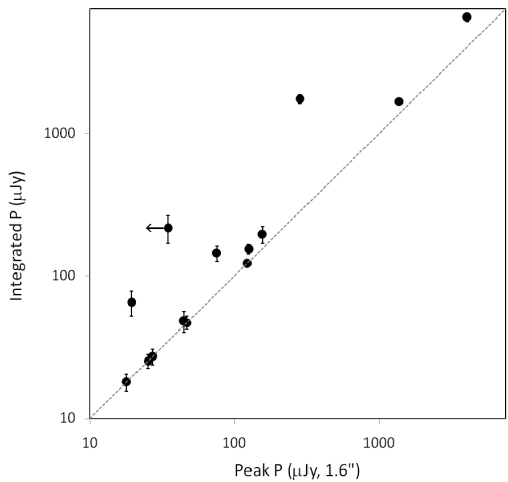}
\end{center}
\caption{\footnotesize{Comparison of integrated polarized flux vs. peak polarized fluxes at 1.6\arcsec~ resolution. Integrated polarized fluxes for the extended sources were derived by integrating over the {\em polarized intensity} images at  10\arcsec~  resolution.}}
\label{compareres}
\end{figure}

For the current observations, we can get a rough look at the effects of resolution by comparing the results of our 1.6\arcsec~ automated finding procedure to the visual inspection we did of extended sources at 10\arcsec~ resolution. The approximate detection limit at 10\arcsec~ is 35-40~$\mu$Jy, only 2-3 times higher than our full resolution results.  However, this assumes that we are searching only a very small area where extended regions are seen in total intensity, and so chances of spurious detection are greatly reduced.  In Figure \ref{compareres} we compare the peak flux at 1.6\arcsec~ to the peak flux at 10\arcsec~ for each extended source.  Note that these peaks are not necessarily from the same location, as in J123538.5+621643 and J123644.3+621132.  One source is {\em only} detected at 10\arcsec~ resolution, J123611.2+622810, while two of our compact sources would fall below the detection threshold at the lower resolution.

A possible bias in our 1.6\arcsec~ number counts comes from searching for polarized flux only at the peak pixel in total intensity.  If the source is slightly resolved, and the peak polarized flux is slightly offset, then it could drop below our detection threshold.  It is not currently possible to quantify this effect, because we do not know the statistical distribution of polarized structures in sources $\sim$1\arcsec~ in size.  We note that we did include one source in our counts,  
J123538.5+621643, even though it rose above our detection threshold at 1-2 pixels off the total intensity peak;  it was discovered during our visual inspection of extended sources. At low resolution, this source shows strong polarization, but not at the position of the high resolution polarized peak (see Figure \ref{xtd2}).

In order to assess whether there were other sources that might have been detectable only at 10\arcsec, such as J123611.2+622810, we searched the total intensity 10\arcsec map and found no additional sources with a peak flux I$>$60$\mu$Jy.  Any sources below this limit
would have required fractional polarizations above 60-70\% to be
detected in polarization and thus contribute to the polarized source counts, and would be a distinct population from anything already detected here.

\subsection{Fractional Polarization Distribution}

We examined the distribution of fractional polarization, $\Pi(I)\equiv\frac{p}{I}$, distributions for comparison with similar studies in the literature, where we assume that $\Pi$ may change as a function of I.  Figure \ref{ivf} shows several different measures of the percent polarization as a function of total intensity both for the GOODS-N field and from the literature for higher fluxes.  $\times$s indicate averages over many sources in a flux bin, while circles represent individual sources.

The solid circles represent the detected sources from the GOODS-N, using the 1.6\arcsec~ data for the compact sources and the hotspot region of J123538.5+621643, and the total integrated polarization and total intensity data for the extended sources.  Upper limits (95\% confidence) are provided for compact sources only, at 1.6\arcsec resolution. Below I=500~$\mu$Jy, where individual source upper limits would be poorly constrained, we calculated upper limits for groups of $\sim$100 sources each, binned by their total intensities. The details of this procedure are described in Section \ref{stackp}.  Upper limits from those bins are indicated with $\times$ symbols.

At the highest fluxes, we show in Figure \ref{ivf} the results from the \cite{mesa} analysis of the NVSS.  These values represent the median percent polarization for all sources in those flux bins.   For the ATLBS sample \citep{atlbs} we included only sources with I$>$10~mJy, in order to avoid biases from spurious detections and a severe Eddington bias \citep{stack}, and rebinned the data in bins of 50 sources.  The plot indicates the median percent polarization in each bin and the range in which 2/3 of the values are located.  For the ELAIS-N1 sample, we use the median value quoted in \cite{elais1}. 

We discuss the implications of the new GOODS-N results in comparison with the higher flux level findings in Section \ref{discussion}.

\subsubsection{GOODS-N upper limits in total intensity bins}
\label{stackp}

Below I=500~$\mu$Jy we first removed  sources with p$>$14.5~$\mu$Jy, leaving a sample of 479 sources, none of which was individually detected in polarization at 1.6\arcsec~ resolution.  We then considered how best to statistically characterize their fractional polarizations. The procedure for a single polarization image over a large field would be relatively straightforward.  One could stack, e.g., small cutout polarization images for each source, and compare that stack to an appropriate control stack, to avoid problems with the Eddington bias \citep{stack}.  The situation would be more complicated in the case where Faraday synthesis is used to search over a range of Faraday depths, and the maximum amplitude is found for each source.  In practice, however,  we re-simplified this problem by picking a single polarized intensity frame from the Faraday cube, namely p$_{source,i}$=$\Phi$(20~\rpm,x$_i$,y$_i$),  corresponding to the RM of the galactic foreground screen.\footnote{p$_{source,i}$ values are normalized to 1.5~GHz, as described earlier.} Therefore, all of our upper limit results apply only to sources with {\em intrinsic} rotation measures of $\sim$0$\pm$53~\rpm, the half-width of the RMSF (Figure \ref{rmsf}). If the distribution of {\em intrinsic} RMs of these non-detected GOODS-N sources were similar to that of the brighter NVSS sources at $|$b$|>$30$\deg$, (GOODS-N is at b=55$\deg$), then our search range would cover 98\%  of the sources.  The alternative would have been to use $\Phi_{max}(x,y)$ instead of $\Phi$(20,x,y), and be sensitive over the full sampled range of $\pm$600~\rpm, which could recover the remaining 2\% of potential sources at the expense of a factor of several in the loss of sensitivity.

  We sorted the 479 sources below I=500~$\mu$Jy that were nondetected in polarization in order of their primary beam corrected total intensities (I$_{pb}$), to create five different flux bins, each with $\sim$96 sources\footnote{One bin had 95 sources.}.  We calculated the median I$_{pb}$ for each bin.   We then recorded the value p$_{source,i}$=$\Phi(20,x,y)$ at the position of each total intensity peak, and p$_{control,i}$=$\Phi$(20,x$\pm$9, y$\pm$9) for four positions each offset by 9 pixels ($\sim$3 beams) in both RA and Dec.   We then calculated the rms value of p$_{source}$ and p$_{control}$ in each of the five flux bins, finding $<p_{source}^2>^{\frac{1}{2}}\sim <p_{control}^2>^{\frac{1}{2}} \sim 5.3~\mu$Jy in each case.  Thus, we found no evidence for excess polarized flux in the binned sources.
\setcounter{table}{2}
\begin{table}[!ht]

\caption{Upper limits at 95\% confidence from total intensity bins}
\begin{center}
 \begin{tabular}{c c}
  \hline
  \hline
  Total Intensity &  Median \\
   $\mu$Jy & Polarization \\
\hline
  229 & $<$~2\% \\
  106 & $<$ ~6\% \\
  70 & $<$~7\% \\
  52 & $<$~8.75\% \\
  40 & $<$~9.5\% \\
\hline
 \end{tabular}
\end{center}
{\em Note that the total intensity bins do} not {\em include the sources already detected; those are small in number and would not affect the median polarization upper limits.}
 \label{stackupper}
\end{table}

\begin{figure*}[!ht]
\begin{center}
\includegraphics[width=12cm, height=9cm]{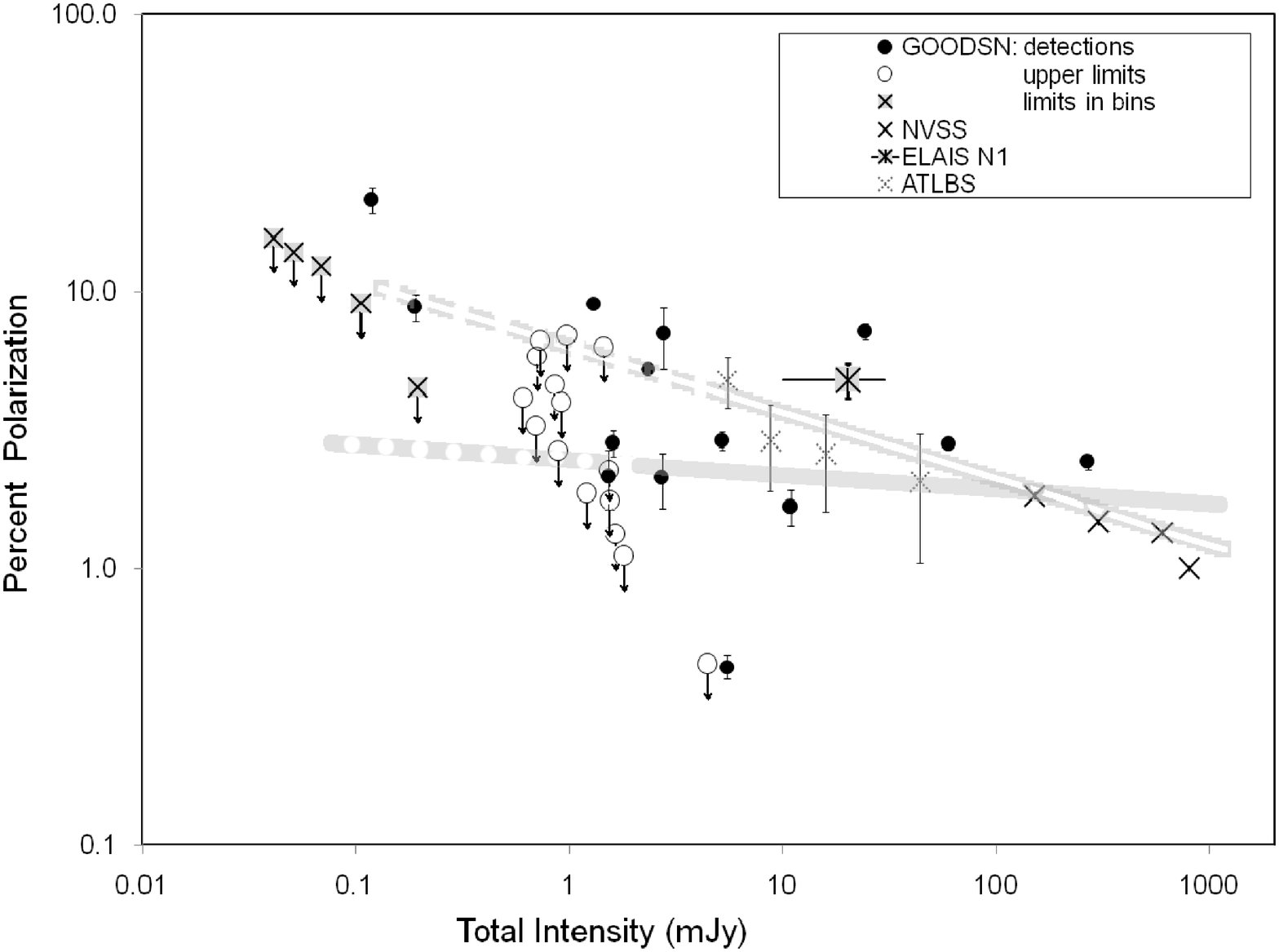}
\end{center}
\caption{\footnotesize{Plot of fractional polarization as a function of total intensity, from current work and from the literature. { There are two types of measurements: Circles designate \emph{individual sources} from the GOODS-N field.  X's represent \emph{binned} results both from the literature and from the GOODS stacking results.  The double grey line represents the previous trends reported {by \cite{atlbs} (using only their data above I=5~mJy) and  \cite{tucci}} (dashed=extrapolated).  The thick grey line represents the recent analysis by \cite{stack} (dashed=extrapolated). }}}
\label{ivf}
\end{figure*}

In order to determine the upper limit to the average fractional polarizations, we inserted a fake signal into the on-source data, and used the KS test to determine at what level the resulting distribution of observed p$_{fake,i}$ became inconsistent at the 95\% level with the control distribution.  Assuming a fixed fractional polarization $\pi_a$, the fake signal was then inserted with a random angle with respect to the observed p$_{source,i}$, as follows:\\
\centerline{p$_{fake,i}$= ( [p$_{source,i}$*cos($\theta_{ran}$)+ $\pi_a$*I$_i$]$^2$}\\
\centerline{ + [p$_{source,i}$*sin($\theta_{ran})$]$^2$ )$^{0.5}$,}\\
where $\theta_{ran}$ is a random angle from -$\pi$ to +$\pi$. This was done separately for each flux bin. 
 The median upper limits at 95\% confidence are then reported in Table \ref{stackupper} and plotted in Figure \ref{ivf}. Such upper limits can be somewhat sensitive to the assumed shape of the underlying fractional polarization distribution \citep{stack};  we assumed the simplest case of a fixed fractional polarization in each flux bin. The calculations could be re-done if there were specific models of interest for the fractional polarization distributions at these low flux levels.

Adding these new measurements and constraints to those in the literature, we can characterize the fractional polarization distribution, $\Pi$(I) in three regimes { :  A. I$>$5~mJy; B. 0.13$<$I$<$5~mJy; and C. I$<$0.13~mJy.}  At the highest total intensities {A.} $\Pi_{median}$(I$>$5~mJy) { has been reported to rise} from $\sim$1\% to $\sim$5\% as the flux density decreases \citep{atlbs,tucci}.  { Note that the \cite{tucci} analysis found a significant trend only for steep-spectrum sources.} The current experiment { on the GOODS-N field} does not have sufficient statistics to verify those trends.   

From our GOODS-N data { in the middle flux regime,} {B,  individual sources show  $\Pi_{median}(\sim$0.5~mJy $<$~I~$<$5~mJy)  $<$2.8\%,  and  the 96 sources in our strongest { stacked} flux bin yield $\Pi$(0.13~mJy$<$~I~$<$~0.5~mJy)~$<~$2\%.  The GOODS data therefore fall below the extrapolation of the trends reported from higher fluxes, as shown in Figure \ref{ivf}. This discrepancy would even be much worse had we included the \cite{atlbs} data below 5~mJy.  The problem comes from their incorrect inclusion of a large number of polarized sources with a signal:noise of order unity, as described earlier.  This forces their median derived polarization to rise sharply as the total intensity decreases, and even the data at 5~mJy may be biased.  A fuller analysis of these problems is presented by \cite{stack}. At the same time, our results in this intermediate intensity regime are completely consistent with the recent  analysis of \cite{stack},  where a model of the fractional polarization distribution is compared with the results of stacking sources in total intensity.}  

{ In the lowest flux regime, C. I$<$ 0.1~mJy,  the upper limits on $\Pi$ are too high to provide any useful constraints.}

\section{DISCUSSION}
\label{discussion}
Polarized radio sources are of interest both as background probes of various foreground screens, as well as the information they can provide about radio source physics and populations.  We discuss the implications of the GOODS-N results in these two areas.

\subsection{Faraday background surveys}

Background RM experiments have been carried out to illuminate the Faraday structure of the Milky Way \citep{taylornvss,simard81}, a few nearby galaxies \citep{han98,gaensler05,mao08} and the lobes of the nearby radio galaxy Centaurus A \citep{fornax}.  They have also been proposed as background probes of supernova remnants, HII regions, galaxies \citep{step08} as well as clusters of galaxies \citep{govoni10,feretti12}.  These background probes are especially useful when there is no diffuse polarized synchrotron emission from the object of interest.  This is true, e.g., for the central regions of galaxy clusters, where new background studies would be free of the biases found in existing studies, which are dominated by polarized sources actually embedded in the clusters \citep{clarke01, bonafede10, govoni06}.  For such samples, it is very difficult to separate Faraday structure associated with the individual radio sources from that of the overall cluster ICM \citep{laing08,guidetti12,rudnick03}. 

For all of these background experiments,  two key properties determine their ability to probe the foreground Faraday screen, viz., the number of background sources and $\sigma_{RM,bkgrd}$, the scatter in RM that the background sources would have {\em in the absence of the foreground screen under investigation}.  $\sigma_{RM,bkgrd}$ arises from contributions local to each source, from their random path through intergalactic space, and from the Milky Way.  One estimate of $\sigma_{RM,bkgrd}$ comes from our analysis of the \cite{taylornvss} RMs from the NVSS survey (Rudnick et al., in prep.). We calculated a residual RM (RM$_{res}$) for each source by subtracting from its RM the median RM of all sources within a circle of 3$^o$ radius.  We found that $<|$RM$_{res}|>$ decreases both with increasing polarized flux and with increasing galactic latitude.  These decreases level off above p$\sim$15~mJy and $|$b$|$=~30$^o$.  Using these cuts, the distribution of $|$RM$_{res}|$ for the remaining 2941 sources is shown in Figure \ref{RMnvss}.  Two thirds of the values have $|$RM$_{res}| <$12.5~\rpm, which provides an  estimate of $\sigma_{RM,bkgrd}$, averaged over  Galactic latitudes above $|$30$^o|$.  This is essentially the same as the rms scatter of 11~\rpm in the GOODS-N sample, after eliminating the three weakest sources.  These global values are only slightly larger than the scatter of $\sim$9-10~\rpm \citep{poleRM1,poleRM2, poleRM3} seen at the North Galactic Pole, and even less than seen at the South Galactic Pole. \cite{poleRM3} model this scatter as coming approximately equally from the Milky Way and from the extragalactic sources themselves.  These various analyses all show that any  background experiment using independent background sources is therefore limited to  $\sigma_{RM,bkgrd}\ge$10~\rpm.  There is likely some contribution to the  scatter in shallow samples from degree-scale structure in our Galaxy, but this should be insignificant for the much smaller GOODS-N field size.

 $\sigma_{RM,bkgrd}$, if calculated as a true rms scatter, can be very sensitive to the presence of outliers, as we have here.  We therefore suggest using the range of RMs containing 2/3 of the distribution as a better estimator of the scatter.  In this GOODS-N survey, we eliminated the outliers on the basis of their low total and polarized intensities. In other experiments, as discussed in Section \ref{highRM},  a scientific judgement would need to be made whether the outliers represent some peculiarity of the background source, or actually indicate an unusual feature in the foreground under study.

As one example of how our new number counts affect RM background experiments, we assume a survey with a 10~$\sigma$ detection threshold of 100~$\mu$Jy, reflecting the 10~$\mu$Jy target sensitivity for the POSSUM Survey using ASKAP \citep{possum} and the WODAN Survey using Westerbork's APERTIF system \citep{rott11}. The high detection threshold reflects the need to search over a wide range of possible Faraday depths;  if the search can be restricted to a narrower Faraday depth range, then the threshold can be reduced, so we also consider a 50$\mu$Jy limit. For each threshold we calculate the number of expected sources, and $\sigma_{tot}$ their expected scatter in RM.  

 If we assume very optimistically that all 10\arcsec~ polarized fluxes are a factor of two higher than at 1.6$\arcsec$ resolution, then we would have a density of 32$\pm$15 (49$\pm$17) sources per square degree at 100$\mu$Jy (50$\mu$Jy).The large errors are due to the small numbers of bright sources in the GOODS-N field. No reduction is made for high $|$RM$|$ sources, since we found none at these high polarized fluxes.  {\em Note that this paragraph is a correction from the published version, for which the correction will appear as an Erratum.}

The approximate measurement error in RM at the survey limit (signal:noise=10) is then  RM$_{err}$= $\frac{1}{2*10}$*(RMSF\_width), where RMSF, the Rotation Measure Spread Function, has a width given by \cite{brentjens} as RMSF\_width~=~2*$\sqrt(3)$/$(\lambda_2^2 - \lambda_1^2)$.  Using the currently planned POSSUM bandwidth of 1100 - 1400 MHz, this yields an RMSF\_width=120~\rpm~ and an RM$_{err}$=6~\rpm~ at the survey limit.  Adding this in quadrature with $\sigma_{RM,bkgrd}$ from above yields a total error $\sigma_{tot}\sim$14~\rpm ~for each measurement. At a 5$\sigma$ threshold of 50$\mu$Jy, the total error goes up to $\sigma_{tot}\sim$17~\rpm.  We assume, for simplicity, that each background source will provide a single RM at a resolution of 10$\arcsec$. 

For the simplest experiment on a cluster of galaxies, to determine the central field strength, we can attempt to measure $<RM^2>^{0.5}$ through the core of the cluster, where the $\beta$ profile leads to a fairly constant density.  Suppose that one expects a signal of order 10$^{2.5}$~\rpm through the core.  Then, since $\sigma_{tot}$ is considerably less than the expected signal, each measurement could provide a statistically significant result.  At an absolute minimum, one would like  $\sim$4-5 background sources seen through the core, just to protect against anomalies and to estimate the RM scatter, which will also depend on the number of field reversals along the line of sight. Given the above number counts, 4-5 sources requires a survey area of $\sim$0.14~square degree at p=100$\mu$Jy,($\sim$0.08~square degree at p=50$\mu$Jy)  equivalent to a circle of radius 13$\arcmin$ (10$\arcmin$).  \cite{core}, summarizing earlier work by \cite{mohr} and \cite{ota}, show that the distribution of cluster core radii r$_c$ is approximately bimodal.  For a characteristic small r$_c\sim$~50~kpc, this requires that the cluster be within a distance of 17(22)~Mpc, which includes only Virgo.  This experiment is therefore not practical for small cores.   For the  larger core radius group, with r$_c\sim$200~kpc,  the cluster could be as far away as 68(88)~Mpc.  The number of X-ray clusters within the corresponding redshifts of 0.016 (0.021) with M$_{500}\ge$10$^{13}$M$_{\odot}$ from the HEASARC Meta-catalog of X-ray clusters \citep{mcxc} is 22(42).  For more massive clusters, with  M$_{500}\ge$10$^{14}$M$_{\odot}$, these numbers drop dramatically to 3(7).   Therefore, an experiment to probe statistically significant samples of even large cluster cores will be challenging for low mass clusters, and impractical for those with higher masses. 

One can adopt a different, perhaps more modest goal for galaxy clusters, e.g., to characterize the scatter in RM as a function of radius, out to perhaps (an extreme value of)  R$_{500}\sim$1~Mpc.  For such a characterization, we assume that no less than 30 sources are desired, to get reasonable sampling in at least several different radial bins.  We assume, again, that $\sigma_{tot}$ will be much smaller than the expected signal, although this is unlikely to be true out to R$_{500}$.  To get 30 background sources, we need an area of $\sim$0.9(0.6)~square degree, or a radius of 32(26)$\arcmin$, at a threshold of 100$\mu$Jy(50$\mu$Jy), respectively.  For 32(26)$\arcmin$ to extend no further than 1~Mpc, the cluster redshift must be $<$0.025(0.03).  Thus, this experiment will be feasible, but again only for the nearest clusters, and with questionable accuracy relative to the expected Faraday signal at the larger radii.

For individual galaxies, \cite{step08} describe a variety of different experiments, the simplest being the ``recognition'' of simple magnetic field structures.  They argue that a few dozen background RMs are sufficient for this task, depending on the contributions of small scale turbulence in the galaxy that could obscure the global patterns.   Assuming that the value $\sigma_{tot} \sim$14\rpm~ is not the limiting factor, then an area of 1(0.6) square degrees for p=100(50)$\mu$Jy is required. If we take an optimistic thermal scale length of 15~kpc (\cite{gomez},  but see discussion in \cite{step08}), then for a face-on spiral, the distance must be less than 1.5(2)~Mpc, restricting such measurements to the local group.  As one comparison, \cite{han98} have already done this experiment on M31, using 18 independent sources with p$>$170$\mu$Jy in $\sim$0.6 square degrees (29 sources per square degree). This is higher than our predicted counts at 170$\mu$Jy, even accounting for the increase due to lower resolution measurements.  The excess is likely due to polarized sources that are actually in M31 or perhaps even small bright patches of diffuse polarized emission.  The individual galaxy experiment may therefore be somewhat more feasible than discussed here, if sources within the galaxy can be properly modeled in terms of their position along the line of sight and any very local effects.
 
The prospects for the SKA are unclear.  Our results show that extrapolating the cumulative number counts from p$>$1~mJy is inappropriate, because the cumulative counts are flattening.  In order to estimate the polarized counts at the target $\sim$1~$\mu$Jy sensitivity level for the SKA, better statistics at the p=10~$\mu$Jy level and below are clearly required.  At sufficiently low levels, polarization from the SF population may start to become visible, and the number counts could again rise more quickly than in 15-500 $\mu$Jy range.  This is critical, if we are to extend the above galaxy or cluster studies beyond the very nearest ones, or to tackle a much more challenging experiment described by \cite{beckg}.  They express scientific interest in mapping of the magnetic field structure in the outer parts of galaxies and note that with $\sigma_{tot}\sim$1\rpm~, fields weaker than $\sim$1$\mu$G in a halo of 10$^{-3}$/cm$^3$ could be detected.  For such experiments, the intrinsic scatter between sources will require averaging over hundreds of sources to reach the desired precision. To reach this density of sources, a major new population of polarized sources would have to emerge.

\subsection{Source populations}

The change in slope of the polarized number count distribution below p$\sim$1~mJy, as well as the end to the trend for increasing fractional polarization at lower flux densities, could in principle be due to either a change in the populations of sources being studied, or changes in the physical properties of a single population.  Although a detailed modeling of these issues is beyond the scope of this work, some comments on our findings are useful.

There is no single pattern of what types of structures are being detected in polarization.  They include hot spot and lobe regions, cores and jets, and compact sources.  Looking at the luminosities of our detected sources (see Table 2), we find  one, J123451.7+620238, with a clear FRII (10$^{26.8}$W/Hz) luminosity \citep{ledlow} and structure.  Almost all the other sources are in the luminosity transition region between FRIs and FRIIs, assuming a redshift of z=1 for the sources with no redshift or unidentified. One source,  J123611.2+622810 has an unusual diffuse morphology, and no obvious association with objects in the SDSS, 2MASS-K, ROSAT broad band, IRAS 12~$\mu$, or SFD surveys.  Another source, a marginal detection,  J123620.7+622510, would be in the starburst range of luminosities at 10$^{22}$ W/Hz.  

The most striking result from this work is the flattening of the { \emph{polarized counts} } below p$\sim$1~mJy {(see Figures 10 and 11)};  above this value, previous work shows an approximately Euclidean result,  $d$log~N($>$p)/$d$log(p) $\sim$ -1.5, while below 1~mJy we find  $d$log~N($>$p)/$d$log(p) $\sim$ -0.6 .  This is generally in line with the expectations of the semi-empirical model of \cite{wilman} and the followup by \cite{osullivan}, where the FRII population begins dropping sharply below 1~mJy.  In total intensity counts, the expected distribution remains almost Euclidean, due to the combination of faint FRI galaxies and the emergence of star-forming populations.  In fact, \cite{morrisonG} find a Euclidean distribution in the range 30~$\mu$Jy~$<$~I~$<$~1~mJy in the GOODS-N field, consistent with other work.   However, in polarization, the star-forming population is not expected to  make a significant contribution to the polarized counts, and the dropoff predicted by \cite{osullivan} is confirmed here, although the total number of sources we find are still significantly lower than their predictions.  Note that our observed dropoff appears inconsistent with counts from the ELAIS-N1 field \citep{elais2} but as we have  previously noted, their completeness corrections are quite high, and uncertain, at the lowest fluxes.

In terms of the fractional polarization distributions, \cite{tucci} summarize the data and the various possible explanations for the rise in median fractional polarization with decreasing total intensity, as observed down to I$\sim$5~mJy.  We have reasonable statistics in the range from $\sim$0.5~$<$~I~$<$~5~mJy, and show that the median polarization percentage is $<$2.8\%  comparable to, or a bit lower, than seen above I$\sim$~5~mJy. Thus, the trend for increasing fractional polarizations does not appear to continue to lower fluxes.  

\begin{figure}[!ht]
\begin{center}
\includegraphics[width=6cm]{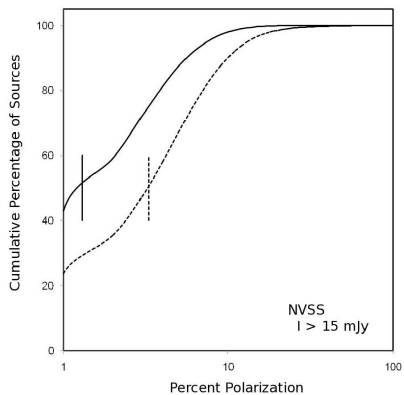}
\end{center}
\caption{\footnotesize{Cumulative distribution of percent polarization in the NVSS, for unresolved (solid line) and resolved (dashed line) sources.  Medians for each type are shown and are 1.3\% and 3.3\%, respectively.{
 Note that the extended source polarizations are uncertain because the NVSS catalog does not report true integrated polarized fluxes.}}}
\label{polsize}
\end{figure}

To explain this behavior with changing source populations \citep{mesa,atlbs} we could invoke increasing dominance of faint polarized FRI sources (the population called radio-quiet-AGNs in \cite{elais1}), and then at lower fluxes, increasing dominance by star-forming galaxies.  Such modeling is beyond the scope of this work, and needs to incorporate the ratios between flat- and steep-spectrum sources and the trend for larger sources to be more polarized \citep{vigotti}. One striking result from the GOODS-N observations is that the polarized sources are dominated by the very small fraction of sources that are quite extended, up to $>$50\arcsec, while the median size is only 1.2\arcsec~ \citep{om08, win03}. This size dependence is even apparent in the NVSS survey, as shown in Figure \ref{polsize}, where we plot the distribution of fractional polarizations (including upper limits) for the $\sim$235000 unresolved sources and for the $\sim$184000 resolved sources, respectively, with total intensities $>$15~mJy.

The issue of changing populations of polarized sources will be addressed by upcoming surveys such as POSSUM, which will have the statistics to further examine possible evolution and environmental effects, although its resolution of 10\arcsec~ will limit the amount of structural information.  Surveys at higher frequencies, e.g., \cite{jackson} are required to isolate the role of depolarization, which will contribute to our understanding of the physical nature of the sources and their interaction with their surroundings.



\section{Conclusions}

Deep observations of the GOODS-N field at 1.4~GHz using the VLA were used to conduct an automated search for polarization from total intensity sources at 1.6\arcsec~ resolution, and a visual search of extended sources at 10\arcsec~ resolution.  

$\bullet$ For the automated search, where the completeness, contamination and available search area as a function of peak polarized flux level are well characterized, we detect a flattening of the cumulative polarized number counts down to 14.5~$\mu$Jy. Above P=1~mJy, other work shows a cumulative slope $d$log(N($>$P))/$d$log(p)$\propto$P$^{-1.4}$, while for P$<$1~mJy, we find $d$log(N($>$P))/$d$log(p)$\propto$P$^{-0.6}$.  Most of the GOODS-N detections are from extended sources with $\sim$0.2$<$z$<$1.9, and are much brighter than the survey detection threshold. 

$\bullet$  The predicted number of polarized sources for upcoming surveys such as POSSUM, with a 5$\sigma$ threshold of 50~$\mu$Jy, is $\sim$35$\pm$10 per square degree, based on the 1.6\arcsec~ counts. Integrated polarized fluxes at 10\arcsec~ resolution may be a factor of two higher, boosting the counts. The rms dispersion in RM, for most of the sources, is 11\rpm~ around the Galactic foreground level;  a small fraction of the sources have high values of $|$RM$|$, making them unsuitable for study of foreground Faraday screens.  At these source densities, individual source foreground studies will only be possible for the nearest galaxies and clusters of galaxies.

$\bullet$ We also find, for 0.5~mJy~$<$~I~$<$~5~mJy, an end of the trend seen at higher fluxes for the fractional polarization to rise with decreasing total intensity.  This, and the number counts, are consistent with the dropoff in the radio galaxy population and the increasing dominance of star-forming sources in the surveys.  For I$<$100$\mu$Jy, we can establish only weak upper limits on the median fractional polarization.

$\bullet$ The detailed distributions of polarized number counts and fractions are dependent on whether peak fluxes or some form of integrated polarized fluxes are used, and on the angular resolution of the survey;  lower resolution surveys can pick up lower surface brightness polarized emission, but can also suffer from depolarization from changes of polarization position angle (intrinsic or RM-caused) within the beam.

\vskip 1in

\begin{acknowledgements}
 The National Radio Astronomy Observatory is a facility of the National Science Foundation operated under cooperative agreement by Associated Universities, Inc..
 Support for the work by LR comes, in part, from NSF grant AST-1211595 to the University of Minnesota.  We thank J. Stil for a number of useful conversations and comments on the manuscript and C. Hales for catching
a calculation error in Section 6.
\end{acknowledgements}

\newpage
\setcounter{table}{0}
\begin{sidewaystable}
\begin{center}

\resizebox{\columnwidth}{!}{
   \begin{tabular}{|c|c|c|c|c|c|c|c|c|c|c|c|}  \hline
SOURCE			&	RA			&	Dec			   &Distance	&P$_{1.6}^{obs}$ & P$_{1.6}^{pb}$ &RM$_{1.6}$&I$_{1.6}^{pb}$		&\%P$_{1.6}$	&P$_{tot}^{pb}$	&I$_{tot}^{pb}$		&\%P$_{tot}$	\\ 
~ & & 		&					(")	&	$\mu$Jy	&	$\mu$Jy	&	\rpm	&	$\mu$Jy	&	~	&	$\mu$Jy	&	$\mu$Jy	&	~	\\ \hline
Compact Sources	&		&   &  & & & & & & & & \\ 
J123426.8+621454	&	12	34	26.80	&	62	14	54.9	&	994	&	47	&	121	&	19(5)	&	2325(70) 	&	5.2(0.4)&	121(4)	&	2325(70)		&	5.2(0.2)	\\
J123555.1+620901	&	12	35	55.14	&	62	09	01.5	&	445	&	15	&	17	&	372(15)	&	189 (8)		&	8.8(2.4)&	7(2)	&	189(8)		&	8.8(1)	\\
J123620.7+622510	&	12	36	20.74	&	62	25	09.8	&	749	&	16	&	25	&	414(14)	&	118(5)		&    21.4(5.5)	&	25(3)	&	118(5)		&	21.4(2.3)	\\
J123718.7+620355	&	12	37	18.73	&	62	03	55.7	&	590	&	18	&	24	&	19(12)	&	5466(165)	&  0.4(0.1)	&	24(2)	&	5466(165)	& 	0.4(0.04)	\\ \hline
Extended Sources	&	&  & & & & & & & &  &\\																							
J123451.5+620246	&	12	34	51.77	&	62	02	38.9	&	1028	&	1453	&	4068	&	26(1)	&	39667(1190)	& 10.3(0.1)	&	6522(420)	& 269190(8075)	&	2.4(0.2)	\\
J125338.1+621032	&	12	35	38.13	&	62	19	32.2	&	637	&	987	&	1382	&	27(1)	&	14943(450)	& 9.2(0.1)	&	1679(70)	& 59782(1795)	&	2.8(0.1)	\\
J123538.5+621643	&	12	35	38.53	&	62	16	42.9	&	510	&	15	&	20	&	175(15)	&	62(4)		& 31.6(6.7)	&	58(13)	&	2716(110)	&	2.1(0.5)	\\
J123550.6+622757	&	12	35	50.64	&	62	27	57.5	&	987	&	60	&	156	&	15(4)	&	3665(110)	&4.3(0.2)	&	181(26)	&	10834(360) 	&	1.7(0.2)	\\
J123611.2+622810	&	12	36	11.22	&	62	28	10.6	&	954	& $<$14.5	& $<$34.8	&	-	&	16(4)		&	-	&	194(48)	&	2768 (195)	&	7.0(1.7)	\\
J123644.3+621132	&	12	36	44.39	&	62	11	33.0	&	94	&	47	&	47	&	41(5)	&	775(25)		&6.1(0.4)	&	45(5)	&	1596(80)	&	2.8(0.3)	\\
J123655.8+615659	&	12	36	55.85	&	61	56	59.0	&	958	&	118	&	283	&	28(2)	&	4857(145)	&5.8(0.2)	&	1739(120)&	24268(755) 	&	7.2(0.5)	\\
J123744.1+621128	&	12	37	25.98	&	62	11	28.5	&	280	&	114	&	125	&	23(2)	&	1017(30)	& 12.3(0.3)	&	150(11)	&	5182(165) 	&	2.9(0.2)	\\
J123820.2+621834	&	12	38	20.23	&	62	18	34.0	&	716	&	28	&	45	&	1(8)	&	557(15)		&8.0(0.9)	&	33(8)	&	1520(70)	&	2.1(0.5)	\\
J123911.8+622216	&	12	39	11.81	&	62	22	16.7	&	1134	&	21	&	76	&	-39(10)	&	142(6)		&	$\sim$53	&	118(18)	&	1300(145)	&	9.1(1.4)	\\ \hline
\end{tabular}
     }
\caption{\footnotesize {Individual detections.~~ RA and Dec are listed for position of the optical identification, if any, as listed in Table 2. Otherwise they refer to the radio source center. 
 For the compact sources, total and 1.6\arcsec~ 
entries (the peak fluxes) are identical.  For the extended sources, I$_{tot}^{obs}$ has been integrated over the 10\arcsec~image.  and P$_{tot}^{obs}$ has been integrated over the 10\arcsec~ images of {\em polarized intensity}, with P$_{tot}^{obs}$ corrected for the noise bias of $\sim$23~$\mu$Jy/beam.  Separate primary beam corrections have been applied to P and I, as described in the text.   Errors of 3\% have been added in quadrature to represent uncertainties in the flux scale.  Noise errors for the peak flux 1.6\arcsec~ polarized intensity measurements are $\sim$3.3~$\mu$Jy. }}
                
\end{center}
\end{sidewaystable}
\clearpage

\begin{table}
\setcounter{table}{1}
\begin{center}
\resizebox{10cm}{!}{
    \begin{tabular}{|c|c|c|c|}
        \hline
        SOURCE            	 & K$_s$(mag)	& Redshift  		& log L $\frac{W}{Hz}$ 	\\   \hline 
      Compact Sources   	&         	& ~     	 	&     		 \\     
        J123426.8+621454    	&20.7 		&  ~    		&  25.1 	\\ 
        J123555.1+620901    	& 20.9		& 1.875   		&    24.6        \\ 
        J123620.7+622510*   	& 18.2  	& 0.18$^a$(0.03) 	&   22.0     \\ 
        J123718.7+620355    	&20.8  	 	&  1.58$^a$    		&   25.9  \\  \hline 
        Extended Sources    	& ~           	& ~       		&            \\ 
        J123451.7+620238    	& 18.7		  & 0.7 (0.09)  		&    26.8   \\ 
        J123538.1+621932    	& 19.8		& 1.222			&   26.7    \\ 
        J123538.5+621643* 	& 18.9$^b$	&  0.72$^a$ 		&   24.8 \\
        J123550.6+622757      	& 17.0  	&   0.504  		&  25.0 \\ 
        J123611.2+622810   	&  		&      			&   25.0    \\ 
	J123644.3+621132    	& 18.7 		&  1.013    		&    24.9   \\ 
	J123655.8+615659    	& 18.3    	&     0.38$^a$(0.06)    &    25.1     \\ 
	J123725.9+621128    	& 20.2 		&   1.641    		&  26.0    \\ 
        J123820.2+621834       & 19.5$\ddagger$	&    0.82$^a$   	&  24.7  \\ 
        J123911.8+622216    	& 17.3   	&    0.34$^a$(0.02)     &   23.7     \\

        
        \hline
    \end{tabular}
}
\end{center}
\caption
 {\footnotesize {
K$_s$  magnitudes from \cite{wang}. 
\vskip 0.01in Redshifts were based on SDSS DR9, except for J123555.1 from 
\cite{chapman}
\vskip 0.01in ~ ~ ~ and for  J123718.7, J123538.5, J123555.1, and J123820.2,  obtained from \cite{rafferty}. 
\vskip 0.01in ~ ~ ~ Redshift errors, when available, given in parentheses.
\vskip 0.01in  For sources with no redshift, z=1 is assumed, and luminosities  calculated assuming a 
\vskip 0.01in ~ ~ ~  flat cosmology with H$_0$=71~km/s/Mpc $\Omega_{matter}$=0.27, and $\Omega_{\Lambda}$=0.73.  
\vskip 0.01in $^a$ photometric redshift;  $^b$ blue stellar object in SDSS};  $\ddagger$ uncertain ID; 
\vskip 0.01in  * = marginal polarization detection at 1.6\arcsec.   }

\end{table}
\end{document}